\documentclass[aps,groupedaddress,fleqn,preprint]{revtex4}
\usepackage{graphicx}
\usepackage{xcolor}
\usepackage{amsmath}
\usepackage{float}
\usepackage{physics}
\usepackage{amsfonts}

\newcommand{\comment}[1]{}
\begin{document}
\title{Transition in the supercritical state of matter: review of experimental evidence}
\author{C. Cockrell$^1$, V. V. Brazhkin$^2$ and K. Trachenko$^1$}
\address{$^1$ School of Physics and Astronomy, Queen Mary University of London, Mile End Road, London, E1 4NS, UK}
\address{$2$ Institute for High Pressure Physics, RAS, 108840, Troitsk, Russia}


\begin{abstract}
A large and mostly unexplored part of the phase diagram lies above the critical point. The supercritical matter was traditionally believed to be physically homogeneous with no discernible differences between liquidlike and gaslike states. More recently, several proposals have been put forward challenging this view, and here we review the history of this research. About a decade ago, it was proposed that the Frenkel line (FL), corresponding to the dynamical transition of particle motion and related thermodynamic and structural transitions, gives a unique and path-independent way to separate the supercritical states into two qualitatively different states and extends to arbitrarily high pressure and temperature on the phase diagram. Here, we review several lines of enquiry that followed. We focus on the experimental evidence of transitions in deeply supercritical Ne, N$_2$, CH$_4$, C$_2$H$_6$, CO$_2$ and H$_2$O at the FL detected by a number of techniques including X-ray, neutron and Raman scattering experiments. 
We subsequently summarise other developments in the field: recent extensions of analysis of dynamics at the FL, quantum simulations, topological and geometrical approaches, the universality of properties at the FL including transport properties, their fundamental bounds and the implications of the supercritical crossover for astrophysics and planetary science. Finally, we review current theoretical understanding of the supercritical state including its thermodynamic theory and list open problems in the field.

\end{abstract}

\maketitle
\tableofcontents

\section{Introduction: liquid, gas and supercritical states}

\subsection{Difference between liquids and gases}

The phase diagram of matter surveys the areas where solid, liquid and gas exist as physically distinct phases separated by transition lines. The solid-liquid melting line is not bound at high temperatures or pressures as long as the system remains chemically unaltered. The two other transition lines are finite in length: the low-lying solid-gas sublimation line terminates at the triple point, and the liquid-gas boiling line which branches from it terminates at the critical point (we do not consider transitions due to quantum effects at low temperature). A vastly larger area of the phase diagram lies beyond the critical point: the supercritical state of matter, the subject of this review.

Until fairly recently, there was no reason to survey matter well above the critical point in any detail. That part of the phase diagram was considered to be unremarkable and physically homogeneous, with no discernible differences between liquid-like and gas-like states \cite{Landau1980,Ma,Hansen2003,Hansen2006,Kiran2000a}. This statement was just about all that was ever mentioned about the supercritical state in textbooks from theoretical point of view. The first textbook explaining that the supercritical state of matter is in fact rich with exciting physics appeared this year (2021) only \cite{Proctor2020a}.

Not far above the critical point, persisting critical anomalies can continue to conditionally separate liquid-like and gas-like states. More recently, experiments, modelling and theory (see, e.g., Refs. \cite{Prescher2017,Smith2017,Proctor2018,Pipich2018,Proctor2019,Cockrell2020a,Pipich2020,Pruteanu2021,Brazhkin2012,Brazhkin2012a,Brazhkin2013,Trachenko2016,Baggioli2020,Wang2017,Wang2019,Cockrell2021}) have given indications that the entire supercritical state may in fact be inhomogeneous and have states with qualitatively different properties. This indication is fundamentally based on a difference between liquids and gases, and for this reason we discuss this difference in detail below.

We first note that differences between crystals and liquids and the differences between crystals and gases are readily apparent on account of symmetry and long-range order in crystals and its absence in liquids and gases \cite{Landau1980,Ma}. This also explains why there is no critical point on either the melting or sublimation line: the changes of symmetry are discontinuous at these lines, at odds with the difference between phases disappearing at the critical point.

A coherent notion of liquid and gas is a relatively recent development in physics and chemistry. Antoine Lavoisier was the first to suggest that all gases can be condensed into liquids \cite{Lavoisier1862}. In Lavoisier's time, vapours were seen as distinct from gases, also called \textit{elastic fluids}, with the former coming from the evaporation of a liquid and the latter being a distinct state of matter (to which the now-obsolete caloric and aether fluids also belonged). Experiments by Lavoisier, Michael Faraday and Humphry Davy \cite{Faraday1823, Faraday1823a}, and others in which gases were liquefied showed that gases (caloric, aether, and the electric fluid notwithstanding) are vapours and are therefore states of ordinary matter rather than a different type altogether.

At first glance to our modern eyes, the properties of liquids and gases, their commonalities, and their differences are well-known. One does not need to be a physicist in order to answer the question ``is there a difference between a liquid and a gas?'' - it is enough to look at full and empty glasses or consult practical advice (see Figure 1). At the same time, important and interesting questions arise if one attempts to introduce rigorous physical criteria of the difference between the liquid and the gas states of matter. We will see how these questions call for the revision of textbook concepts such as the indistinguishability of the states of matter above the critical point and are also related to the physical origin of the main difficulty which impedes a theoretical understanding of liquids.

\begin{figure}
\begin{center}
{\scalebox{0.45}{\includegraphics[width=\columnwidth]{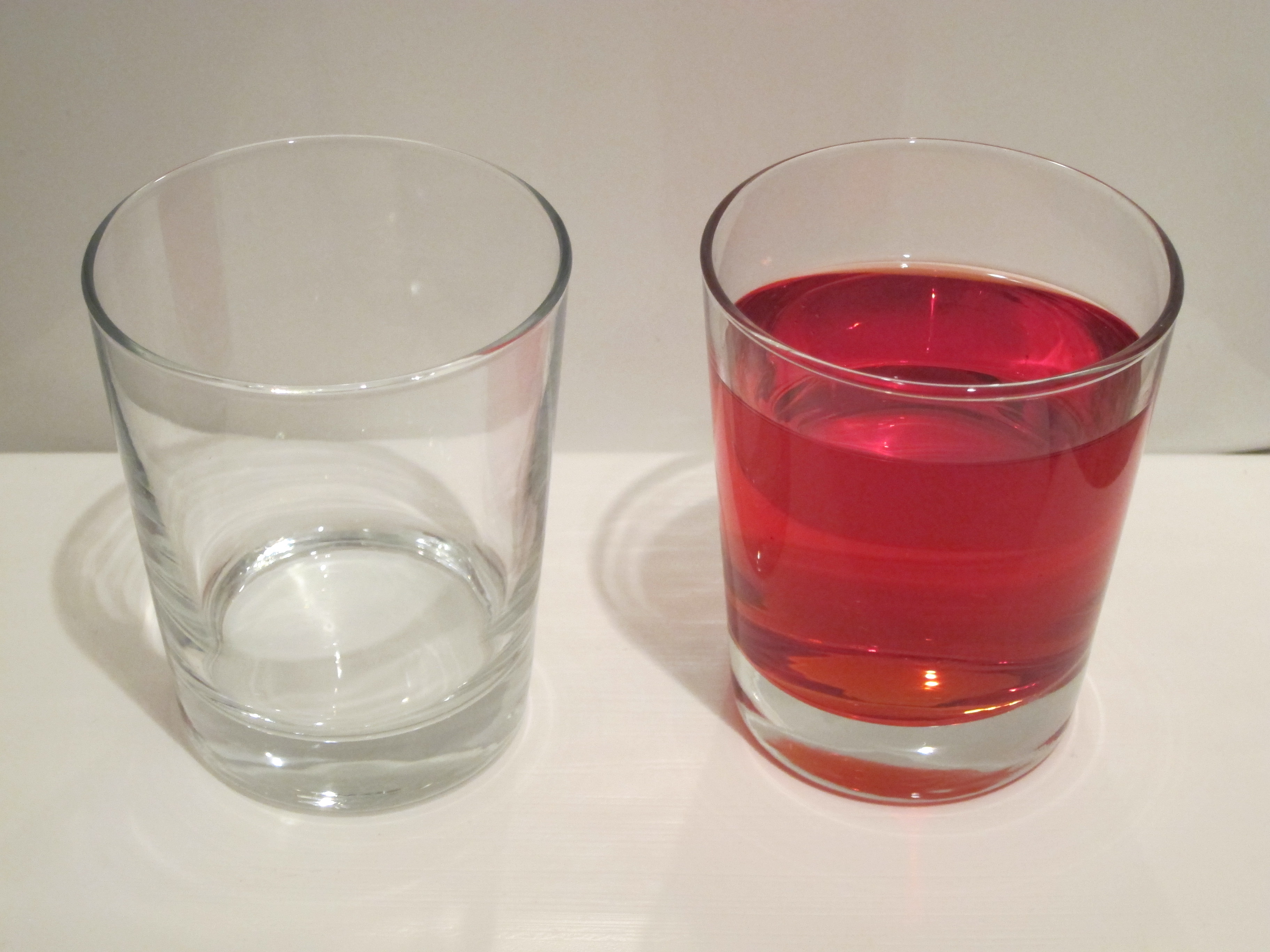}}}
\hspace{0.2cm}
{\scalebox{0.45}{\includegraphics[width=\columnwidth]{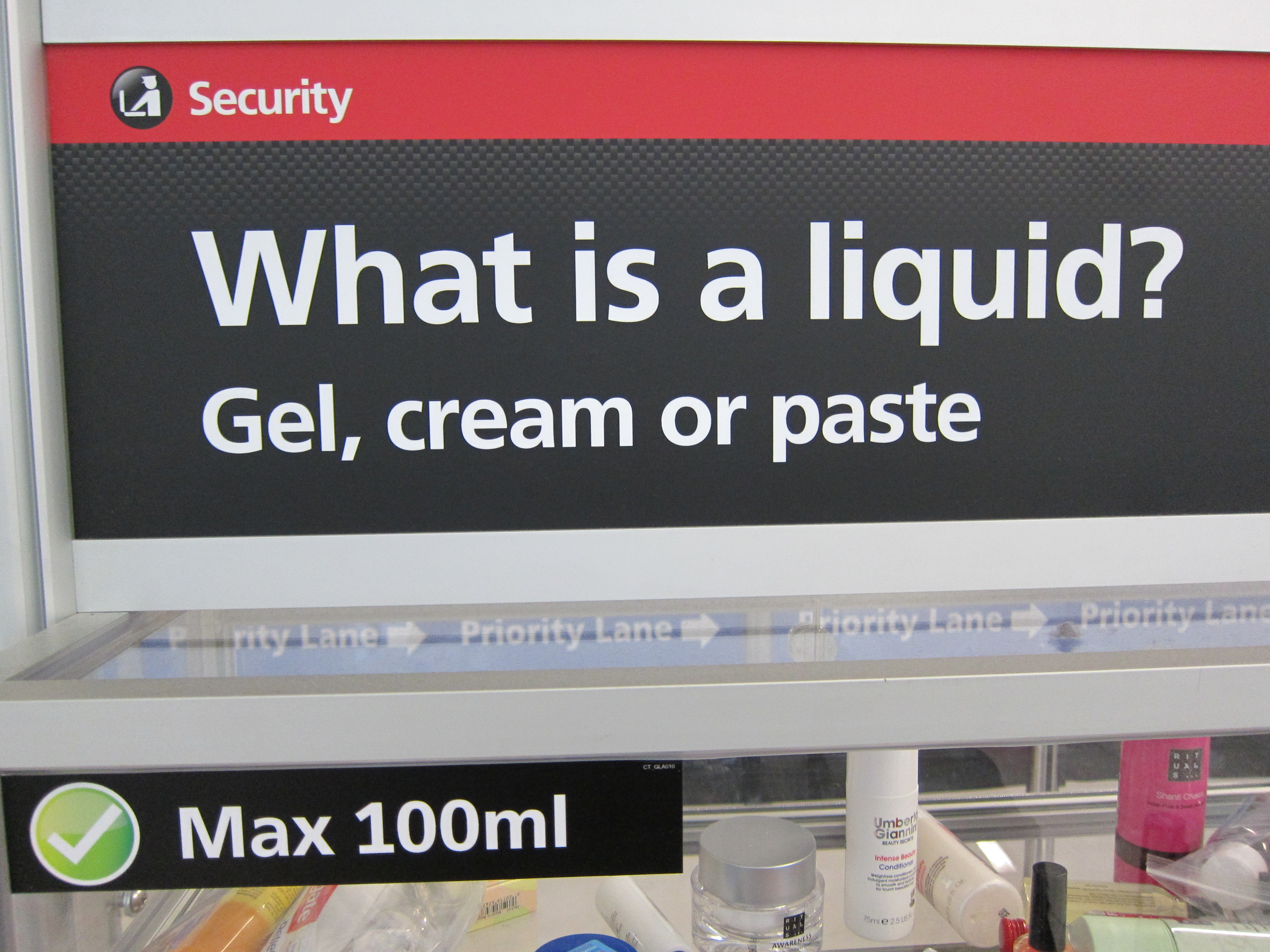}}}
\end{center}
\caption{One does not hesitate to declare the glass to be full of a gas and the glass to be full of a liquid. In more subtle cases, displays such as that partially shown in the second image are intended to help airline passengers who have liquids in their carry-on baggage. Reproduced from V. V. Brazhkin and K. Trachenko, Physics Today 65(11), 68 (2012), with the permission of the American Institute of Physics (https://physicstoday.scitation.org/doi/10.1063/PT.3.1796).
}
\label{6}
\end{figure}

The differences between liquids and gases that are often highlighted first are quantitative. Gases, as a rule, have densities that many (thousands) times smaller than liquids and, consequently, much higher compressibility. Other physical properties of gases and liquids are very different as a result, including viscosity, refractive index and so on. However, rigorous definitions of the states of matter cannot involve quantitative differences only. Indeed, the actual values of density and other properties strongly depend on external parameters: temperature and pressure. As the critical point is approached, the density jump at the liquid-gas transition goes to zero, and the quantitative differences become small. Therefore, the key differences between liquids and gases, from the physical point of view, must be qualitative.

Most textbooks, including modern textbooks dedicated to liquids \cite{Hansen2003}, state two main properties that are different in liquids and gases. The first one is the existence of the free surface in the liquid (the inter-phase boundary with the liquid environment). The second and main property, partially related to the first one, is that a liquid is a condensed state of matter and, consequently, possesses cohesion \cite{Hansen2003}. Stated differently, a liquid has a finite cohesion energy, and the minimum of the free energy is achieved at some finite volume. This implies that a liquid has the finite volume in a certain pressure range, including at zero and negative pressure in the metastable state.

In passing, we note that some textbooks formulate the latter condition in less rigorous terms: a liquid occupies a finite volume in a vessel whereas a gas occupies all volume available to it so that the volume of a gas becomes infinite at zero pressure and finite temperature. However, this criterion has issues. A liquid can only be a stable phase at pressure above the pressure of the triple point (see Figure 2). At smaller pressure including zero pressure, a liquid boils and becomes a gas (which also occupies a finite volume at a finite pressure). Therefore, an experimental verification of whether a system occupies a finite volume and remains in the liquid state depends on the value of pressure above the surface. Our perception of water in a glass being a liquid is only due to the fact that the pressure of water's triple point is 160 times smaller than the atmospheric pressure, and we assume that water would be stable at any value of pressure above the surface. On the other hand, the pressure at the triple point for CO$_2$ is 5 times larger than the atmospheric pressure and, therefore, liquid CO$_2$ at atmospheric pressure does not exist. Yet at higher pressure, for example at 20 atmospheres and -20 C or at 70 atmospheres and +20 C, liquid CO$_2$ can be poured in a glass just as water can. Hence, the difference between a liquid and a gas based on the ability to occupy a finite volume depends on the value of an external macroscopic parameter, pressure. Still, the liquid-gas transition is a first-order phase transition, and metastable states exist on both sides of the transition. This implies that any liquid can exist, for a certain amount of time, in its metastable state on the left of the liquid-gas line in Figure 2, including at zero and even at negative pressure. As a result, we can experimentally verify that the system is the condensed liquid at zero pressure, but only if we succeed to maintain a liquid in the metastable state for a certain amount of time. However, there is a spinodal line for the liquid-gas transition that corresponds to compressibility becoming infinite. The liquid state becomes unstable at the spinodal, and outside the spinodal the liquid can not exist even in the metastable state. Therefore, to experimentally satisfy ourselves that matter at arbitrary pressure is in the liquid but not in the gas state, we have to be in the metastable region of the phase diagram which, furthermore, exists only in a limited temperature range.

\begin{figure}
\begin{center}
{\scalebox{0.5}{\includegraphics{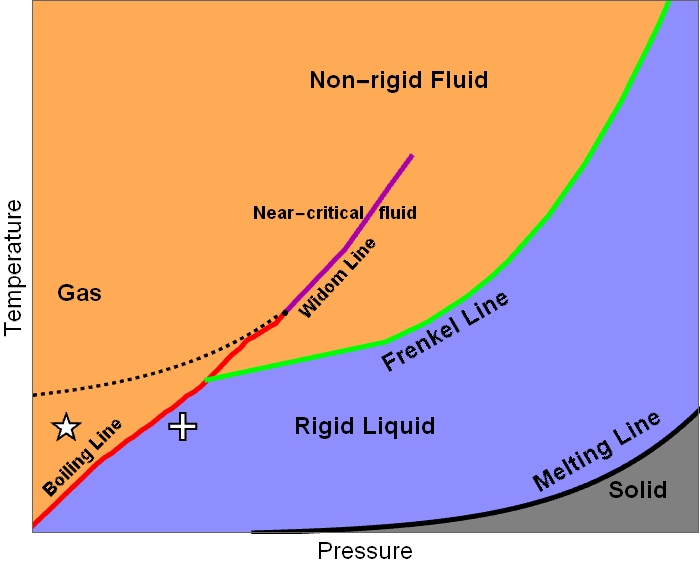}}}
\end{center}
\caption{The phase diagram of matter in the high-temperature and high-pressure range. The boiling line terminates at the critical point, out of which the Widom line and related lines emerge, representing near-critical anomalies of different thermodynamic quantities. The Frenkel line terminates at the boiling line but extends arbitrarily far into the supercritical state, separating rigid liquid-like and nonrigid gaslike fluids. The star and the cross illustrate the areas where the liquid exists in a stable and metastable state, respectively. The metastable state is below the spinodal shown as a dashed line.}
\label{fig:phasediagram}
\end{figure}

A more constructive distinction between liquidlike and gaslike states can be made on the basis of microscopic particle dynamics, in the sense that this property enables us to discuss and predict distinct measurable properties in both states which, importantly, extend above the critical point. This distinction is qualitative rather than quantitative, and can be made at {\it any} value of external parameters such as pressure and temperature, making it free from issues mentioned above.

The key physical process is the qualitative difference between the types of atomic trajectories in liquids and gases. In gases, particles move in straight lines between the collisions after which they change course. In liquids, atomic motion consists of two types: quasi-harmonic vibration around equilibrium positions and diffusive jumps between two neighbouring equilibrium positions \cite{Frenkel1955}. Therefore, atomic motion in a liquid combines both elements of small-amplitude vibrations as in a solid and diffusive motion as in a gas. On temperature increase (or pressure decrease), a particle spends less time vibrating and more time diffusing. The ($P$,$T$) conditions at which the solid-like oscillating component of motion disappears, leaving only the diffusive motion, correspond to the qualitative change of particle dynamics and to the transition of the liquidlike into the gaslike state \cite{Brazhkin2013,Brazhkin2012,Brazhkin2012a}.

The advantage of operating in terms of particle dynamics is that it enables us to delineate a supercritical fluid into two dynamically different states as discussed below. Sufficiently below the critical point, the qualitative change of particle dynamics takes place abruptly during the phase transition when crossing the boiling line with an accompanying large density change. This phase transition and large density jump stop operating above the critical point, however - and this an important point - the qualitative change of particle dynamics can still be discerned. This brings us to the discussion of the Frenkel line in the supercritical state in the next section.

\subsection{The supercritical state}

In 1822, well before many gases had been liquefied and firmly proven to be vapours, the phase diagram received another amendment at the hands of Charles Cagniard de la Tour. In his experiments on acoustics, Baron de la Tour discovered the critical point \cite{Cagniard1822, Cagniard1823} by heating liquid-vapour mixes to the point where no apparent distinction between them was observed. Our view of this state, now called the supercritical state (a name given to it by Thomas Andrews, who elucidated much of its nature \cite{Andrews1869}), has not changed in the 200 years since its discovery. Near critical anomalies behave like extensions to the boiling line, but do not persist deep into the supercritical state which is otherwise seen as homogeneous, unremarkable and lacking in transitions \cite{Landau1980,Ma,Hansen2003,Hansen2006,Kiran2000a}.

Several proposals discussed the physical basis for separating liquid-like and gas-like states not far above the critical point. For example, this included the formation of ``ridges'' corresponding to the maxima of density fluctuations \cite{Nishikawa2000,Nakayama2000,Nishikawa2003,Sato2008}. A later proposal introduced the Widom line, the line of heat capacity maxima persisting above the critical point \cite{Xu2005} (see Fig. \ref{fig:phasediagram}). One can similarly define lines of maxima of other properties such as compressibility, thermal expansion and density fluctuations. These lines, as well as the Widom line, are close to each other in the vicinity of the critical point but diverge away from the critical point \cite{Brazhkin2011,Proctor2020a}. The are two further important properties of these lines. First, the lines of maxima depend on the path taken on the phase diagram and are different on isotherms, isobars, isochores and so on. Second, the maxima of the heat capacity, compressibility, thermal expansion and so on disappear around (2.5-3)$T_c$, where $T_c$ is the critical temperature, and so do the corresponding lines including the Widom line \cite{Brazhkin2011,Proctor2020a}.

More recently and about ten years ago, the Frenkel line (FL) was introduced \cite{Brazhkin2012,Brazhkin2012a,Brazhkin2013} as a dynamical crossover line in the supercritical state. The crossover is in the sense discussed in the previous section and separates the combined oscillatory and diffusive motion below the line from purely diffusive above the line. The FL is illustrated in Fig. \ref{fig:phasediagram}.

The disappearance of the oscillatory component of motion can be discussed on the basis of liquid relaxation time $\tau$, time between two large diffusive molecular jumps in the liquid \cite{Frenkel1955}. $\tau$ defines time a molecule spends vibrating as in solid before jumping to a neighbouring quasi-equilibrium position, and varies in a wide range: from about 0.1 ps at high temperature to 10$^3$ s at the liquid-glass transition. When $\tau$ starts approaching $\tau_{\rm D}$, where $\tau_{\rm{D}}$ is the minimum (Debye) period of oscillation in the system, the oscillatory component of motion is lost. Hence, $\tau\approx\tau_{\rm D}$ is an approximate qualitative criterion of the FL.

There are two quantitative criteria of the FL which give the same line on the phase diagram. The first, dynamical, criterion is based on the behavior of the velocity autocorrelation function (VAF). We note that calculating and discussing VAF in liquids dates back to an early era of computer modelling \cite{Rahman1964}. Oscillatory behavior and monotonic decay of VAF was earlier linked to liquidlike and gaslike behavior respectively \cite{Hiwatari1974,Endo1982}. In similar spirit, we have defined the FL as the line where solid-like oscillatory behavior of the VAF and its minima disappear \cite{Brazhkin2013}. The second quantitative criterion of the FL is thermodynamic and defines the line where the constant-volume heat capacity per atom is $c_V=2k_{\rm B}$. As explained in section IIIc, this value corresponds to two transverse modes disappearing from the system's spectrum because the oscillatory component of particle motion is lost at the FL \cite{Trachenko2016,Brazhkin2013,Kryuchkov2019}. This criterion applies to harmonic or quasi-harmonic excitations where the energy of each mode is $k_{\rm B}T$ but not to idealised and exotic systems such soft-sphere systems with large repulsion exponent or the hard-sphere system as its limit \cite{Brazhkin2012,Brazhkin2013}. We will elaborate on the theory underpinning the importance of $c_V=2k_{\rm B}$ in section IIIB below.

That the system loses shear resistance at all frequencies at the FL is an important point because the ability of liquids to flow is often associated with their zero rigidity that markedly distinguishes liquids from solids. However, this implies zero rigidity at small frequency or wavevectors only, whereas at higher frequency $\omega$ and wavevector $k$, a liquid supports shear stress \cite{Trachenko2016}. On the other hand, the Frenkel line corresponds to the complete loss of shear resistance, at all frequencies available in the liquid, with the result that above the Frenkel line the system can not sustain rigidity at any frequency, becoming, therefore, a gas-like state. This is accompanied by the reduction of constant-volume specific heat to 2$k_{\rm B}$ per particle, corresponding to potential energy of shear modes becoming zero \cite{Trachenko2016,Trachenko2008}. 

The crossover of particle dynamics affects other physical quantities of the supercritical system. Viscosity, speed of sound and thermal conductivity all decrease with temperature in the liquidlike regime but increase with temperature above the line as in gases, i.e. have minima. The diffusion constant crosses over from exponential temperature dependence in the liquidlike regime to power-law dependence as in gases. Depending on the path chosen on the phase diagram, the crossovers and minima of these properties may deviate from the FL but often remain close to it \cite{Brazhkin2012b,Fomin2018path}.

Differently from the Widom line, the FL extends to arbitrarily high pressure and temperature (as long as the system remains chemically unaltered; a chemical and/or structural transformation may result in a new and different FL). Another important difference from the Widom line is that the FL is physically unrelated to the critical point and exists in systems such as soft-sphere system without a critical point or a boiling line \cite{Brazhkin2013}.

In systems with the boiling line and critical point, the FL continues below the critical point and touches the boiling line slightly below the critical temperature \cite{Brazhkin2013}. Crossing the FL by increasing the temperature at a fixed pressure below the critical point corresponds to the transition between gas-like and liquid-like dynamics. Crossing the boiling line on further temperature increase corresponds to a phase transition between two gas-like states with different densities, where the  lower temperature phase is regarded as a dense gas because a cohesive liquid state ceases to exist close to the critical point \cite{Stishov1993}.

Close to the critical point itself, particle dynamics is gaslike in the sense discussed above because the FL touches the boiling line below the critical temperature as just mentioned. However, this is inconsequential for system properties near the critical point because these properties are largely governed by large critical fluctuations and anomalies which set the thermodynamic, dynamical and elastic behavior of the system \cite{Ma,Sengers1986,Anisimov2011}.

In this regard, it is interesting to recall the Fisher-Widom line (FWL) which, similarly to the FL, touches the boiling line slightly below the critical temperature \cite{Proctor2020a}. The FWL separates the fluid state where the radial distribution function (RDF) has an oscillatory decay from the state where the RDF has one single decaying peak only. Experimentally, the FWL has not been detected \cite{Proctor2020a}. Originally deduced for one-dimensional system \cite{fwl}, the FWL was later deduced in a three-dimensional system on the basis of modelling where the truncation of interatomic potential at short distance was found essential to obtain the transition between oscillatory and decaying behavior of the RDF \cite{fwl2,fwl1}. It is currently unclear how far the FWL extends above the critical point \cite{Proctor2020a}.

We note that the three lines mentioned so far, the Widom line, the Frenkel line and the Fisher-Widom line, do not correspond to a first- or second-order phase transition. The terms ``liquidlike'' and ``gaslike'' states we use in relation to the Frenkel line are related to the character of particle motion as described above. However, the FL might be accompanied by a higher-order phase transition, as discussed in section IIIC below.


Finally, the consideration of the different states of matter from the point of view of particle dynamics highlights the origin of the problem of theoretical description of liquids and superciritcal fluids (see Ref. \cite{Trachenko2016} for a detailed discussion). From the point of view of dynamics, only solids and gases are ``pure'' states of matter. In solids, crystals and glasses particle dynamics are purely oscillatory. In gases, the dynamics are purely diffusive and collisional. Physically, this is due to the kinetic energy of particles being much smaller than the energy barriers between various potential minima in solids. In gases, it is the other way around. On the other hand, liquids and supercritical fluids are not a ``pure'' state of matter from the point of view of particle dynamics, but is a ``mixed'' state: it involves both oscillations and diffusive motions, with their respective contributions gradually changing in response of external parameters. It is this mixed state that has been ultimately responsible for the difficulty in constructing a theory of liquids and supercritical fluids \cite{Trachenko2016}.

\subsection{Aims, scope and timing of this review}

The main aim of this review is to compile and discuss the experimental evidence of the Frenkel line. The FL was proposed about ten years ago \cite{Brazhkin2012,Brazhkin2012a,Brazhkin2013}, and the first experimental evidence appeared several years later \cite{Prescher2017}. Significantly more experimental evidence has accumulated since (see, e.g., Refs. \cite{Smith2017,Proctor2018,Pipich2018,Proctor2019,Cockrell2020a,Pipich2020,Pruteanu2021}) involving different X-ray, neutron and Raman scattering techniques and instruments and several important supercritical fluids. Below we review the experimental data in deeply supercritical Ne, N$_2$, CH$_4$, C$_2$H$_6$, CO$_2$ and H$_2$O showing crossovers at the FL. We subsequently summarise other developments in the field, including recent extensions of analysis of dynamics at the FL, quantum simulations, topological and geometrical approaches, the universality at the FL including transport properties and implications for astrophysics and planetary science. Finally, we review current theoretical understanding of the supercritical state and list open problems in the field.

Importantly, the supercritical experiments in Ne, N$_2$, CH$_4$, C$_2$H$_6$ and CO$_2$ discussed below were stimulated and guided by the theory: the experiments followed the state points of the FL mapped in preceding calculations (see, e.g., Refs. \cite{Brazhkin2013,Yang2015}). This is
different from other areas which are experimentally led and where theories follow. Supercritical water is an exception in our experimental list: our discussion of supercritical H$_2$O is based on results published before the FL.

Apart from fundamental understanding of the phase diagram and its supercritical part, recent developments come at an important time from the applications point of view. Supercritical fluids are increasingly employed in a wide variety of applications in industry \cite{Kiran2000a,McHardy1998,Brunner2010,Alekseev2020,Proctor2020a}. Supercritical CO$_2$ and H$_2$O are used for chemical extraction (e.g. producing decaffeinated coffee), biomass decomposition, dry cleaning, fluid chromatography, chemical reactions, impregnation and dyeing, supercritical drying, pharmacology, micro-particle formation, power generation, biodiesel production, carbon capture and storage, refrigeration and using as antimicrobial agents. The environment is one area where the application of supercritical fluids is particularly important. There is a growing pressure from increasing amounts of organic and toxic wastes, and the needs for remediation and effective solutions for waste processing are pressing. The wide range of waste includes toxic chemicals such as biphenyls, dioxins, pharmaceutical wastes, highly contaminated sludge from the paper industry, bacteria as well as stockpiles of warfare agents. Current technologies such as land-filling and incineration have problems related to environmental pollution that are becoming more difficult and costly to handle due to higher treatment standards and limitations. Supercritical liquids have been deployed as effective and environmentally-friendly cleaning and dissolving systems. Destroying hazardous wastes using supercritical water has been shown to be a viable industrial method, and potentially superior to existing conventional methods \cite{Kiran2000a,McHardy1998,Brunner2010}.

High extracting and dissolving efficiency of supercritical fluids is due to the combination of two factors. First, the density of supercritical fluids at typical industrial pressures and temperatures is much higher than that of gases, and is close to that of sub-critical liquids. High density ensures efficient interaction of the molecules of the supercritical fluid and the solute. Second, the diffusion coefficients in a supercritical fluid at typical industrial conditions are 10--100 times higher than for sub-critical liquids, promoting solubility and dissolving. Hence, supercritical fluids have the best of both worlds: high density of liquids and large diffusion constants.

Despite widening industrial use of supercritical fluids, it was noted that their understanding has been lacking \cite{Kiran2000a,McHardy1998,Brunner2010}. Instead, the use of supercritical fluids has been based on empirical observations and test results. It has been widely noted that improving fundamental knowledge of the supercritical state is important for scaling up, widening, and increasing the reliability of these applications (see, e.g., Refs \cite{Kiran2000a,Leitner2002,Eckert1996,Savage1995,Beckman2004,Jessop1999}). In this sense, an earlier observation and explanation that the solubility maxima (``ridges'') lie close to the FL \cite{Yang2015} is industrially relevant. Our review of experimental results is therefore intended to be useful for the industrial use of supercritical fluids.

\section{Experimental evidence in supercritical fluids}

\subsection{Inter-relations between dynamical, thermodynamic and structural crossovers}

In the next several sections, we discuss the experimental evidence of structural correlations in several important supercritical fluids using X-ray, neutron and Raman scattering techniques. Some of this evidence involves our own work. We first recall that the two definitions of the FL above are dynamical, based on the VAF, and thermodynamic, based on $c_V$. A crossover of $c_V$ was detected in molecular dynamics simulations using several fairly sensitive analysis methods \cite{Wang2019}. On the other hand, experimental data of $c_V$ is often based on extrapolation of the equation of state and interpolating functions (see, e.g., Ref. \cite{nist}) and are not of sufficiently detailed quality to study the potentially subtle transition of thermodynamic functions at the FL. Similarly, inelastic scattering experiments revealing dynamics in supercritical fluids such as phonons are challenging, scarce and can be affected by fitting uncertainties. On the other hand, structural studies are easier in comparison. Below we show that the structural crossover is necessitated by the dynamical and thermodynamic crossover at the FL.

The structural crossover is expected on general grounds. As discussed earlier, particles oscillate around quasi-equilibrium positions and occasionally jump between them below the FL. The average time between jumps is given by liquid relaxation time, $\tau$. Therefore, a static structure exists during $\tau$ for a large number of particles below the FL, giving rise to the well-defined medium-range order comparable to that existing in disordered solids \cite{Zaccone2020}. Above the FL, the particles lose the oscillatory component of motion and start to move in a purely diffusive manner as in gases. This implies that the features of the pair distribution function $g(r)$ and its Fourier transform, the structure factor $S(q)$, are expected to be gas-like. As a result, the features of $g(r)$ and $S(q)$ are expected to have different temperature dependence below and above the FL.

The inter-relation between dynamics, thermodynamics and structure can be formalised by writing the system energy as

\begin{equation}
E=E_0+E_T
\label{toten}
\end{equation}

\noindent where $E_0$ is liquid energy at zero temperature and represents temperature-independent background contribution due to the interaction energy and $E_T$ is the energy of thermal motion.

At the same time, the energy per particle in a system with pair-wise interactions is

\begin{equation}
E=\frac{3}{2}k_{\rm B}T+4\pi n \int\limits_0^{\infty} r^2 U(r)g(r)dr
\label{ene}
\end{equation}

\noindent where $n=N/V$ is the concentration (number density), $U(r)$ is the interparticular pair potential, and $g(r)$ is radial distribution function.

If the system energy undergoes the crossover at the FL where $c_V=2k_{\rm B}$, Eq. (\ref{ene}) implies that $g(r)$ should also undergo a crossover. Therefore, the structural crossover follows from the crossover of thermodynamic properties.

Further interrelations between dynamics, thermodynamics and structure follow from writing the thermal fluid energy $E_T$ below and above the line on the basis of collecting excitations, phonons. As discussed in chapter IIIC, this energy has different functional forms and temperature dependence below and above the FL. Below the line, $E_T$ is due to the decreasing number of propagating transverse phonons \cite{Trachenko2016}. These phonons disappear at the FL as discussed above, corresponding to $c_V=2k_{\rm B}$ in the harmonic case. Above the line, $E_T$ is due to the decreasing number of the longitudinal phonons. These phonons become the usual sound at high temperature or low pressure, and its energy tends to 0 in the ideal-gas limit \cite{Trachenko2016}. Different temperature dependence of the phonon energy $E_T$ below and above the FL, together with equating Eqs. \eqref{toten} and \eqref{ene}, implies the crossover of $g(r)$ and structural features.

\subsection{Neon}

We begin with experimental observations in supercritical neon \cite{Prescher2017}. Noble elements have been frequently used as sample cases for liquid theories due to their simplicity \cite{Hansen2003,Balucani1994}. Here too, observations on supercritical noble fluids can be very insightful due to their similarity to the harmonic monoatomic fluid in which the Frenkel picture was conceived.

Samples of supercritical neon underwent x-ray scattering (XS) experiments at a temperature of 290K and pressures of 0.05-3.7 GPa. Neon's critical point is 44.5 K and 2.68 MPa, so these experimental state points are very deeply supercritical and lie well beyond the Widom line.

The structure factor $S(k)$ was extracted from experimental results, and Fourier transformed to yield the pair distribution function (PDF) $g(r)$ \cite{Keen2001}:
\begin{equation}
    g(r) = 1 + \frac{1}{n (2 \pi)^3} \int_0^{\infty} \mathrm{d} k \ 4 \pi k^2 \left( S(k) - 1 \right) \frac{\sin(kr)}{kr},
\end{equation}
where $n = N/V$ is the concentration. It is in these quantities that the crossover at the FL expresses itself - we present the raw data, no fitting or models are necessary. On the basis of molecular dynamics (MD) modelling, the FL in neon was predicted to be at a pressure of 0.6-0.7 GPa at 290 K using the VAF and thermodynamic criteria \cite{Prescher2017}.

\begin{figure}
\begin{center}
{\scalebox{0.4}{\includegraphics{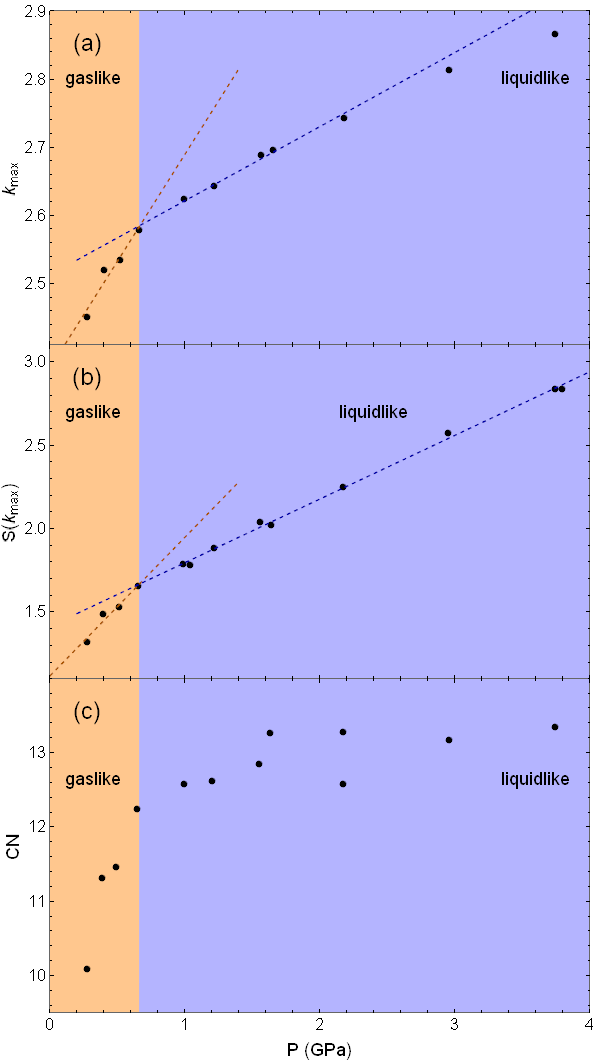}}}
\end{center}
\caption{Structural data of supercritical neon at 290 K: (a) first peak position $k_{\rm{max}}$ of structure factor $S(k)$ as a function of pressure $P$; (b) first peak height; (c) coordination number $CN$ found by integrating $g(r)$ as a function of pressure. The shaded ``gaslike" and ``liquidlike" regions meet at the Frenkel line. The data are from Ref. \cite{Prescher2017}.}
\label{fig:neonplots}
\end{figure}

We first discuss the elements of the structure factor, plotted in Fig. \ref{fig:neonplots}a-b. The first peak position (in $k$) and first peak height each undergo a sharp crossover at $P \approx 0.65$ GPa. Above this pressure, the medium range order evolves much more gradually with changing pressure than below. This is the expected effect of a crossover from liquidlike structure to gaslike, and is centered on the FL predicted by MD simulations of 0.6-0.7 GPa.

The mean occupation of the first coordination shell, the coordination number $CN$, is calculated as
\begin{equation}
\label{eqn:cndef}
    CN = 4 \pi n \int_0^{R} \mathrm{d} r \ r^2 g(r),
\end{equation}
where $R$ is the position of the first minimum in $g(r)$. This as a function of pressure is plotted in Fig. \ref{fig:neonplots}c, where we again observe a crossover at the FL. This crossover is between the closely packed liquidlike regime, which much like $S(k)$ varies weakly under pressure changes and the gaslike regime where coordination depends more directly on density and therefore varies much more strongly with pressure.

These crossovers observed in the simple system of neon set a trend for structural analysis of the FL: the changes observed in $S(k)$ and $g(r)$ describe the \textit{short and medium range order} of the fluid.
The local environment of an average molecule is seen to be sensitive to the liquidlike or gaslike details of the fluid molecular dynamics.

\subsection{Nitrogen}

We first discussed a weakly interacting, monoatomic fluid, and observe modest changes in structural properties across the FL. We move on next to the slightly more complicated molecular nitrogen, N$_2$, the main component of Earth's atmosphere, wherein effects become more pronounced \cite{Proctor2019}. The presence of internal degrees of freedom provides us with additional properties to probe across the FL. Intra and intermolecular dynamics and structure are all linked, which motivated the work in Ref. \cite{Proctor2019}. The authors set out to determine ``what parameters change, and what parameters do not change, when the Frenkel line is crossed". These results are experimental studies prompted by the FL theory (AIMD simulations play only an independent and supporting role).

\begin{figure}
\begin{center}
{\scalebox{0.4}{\includegraphics{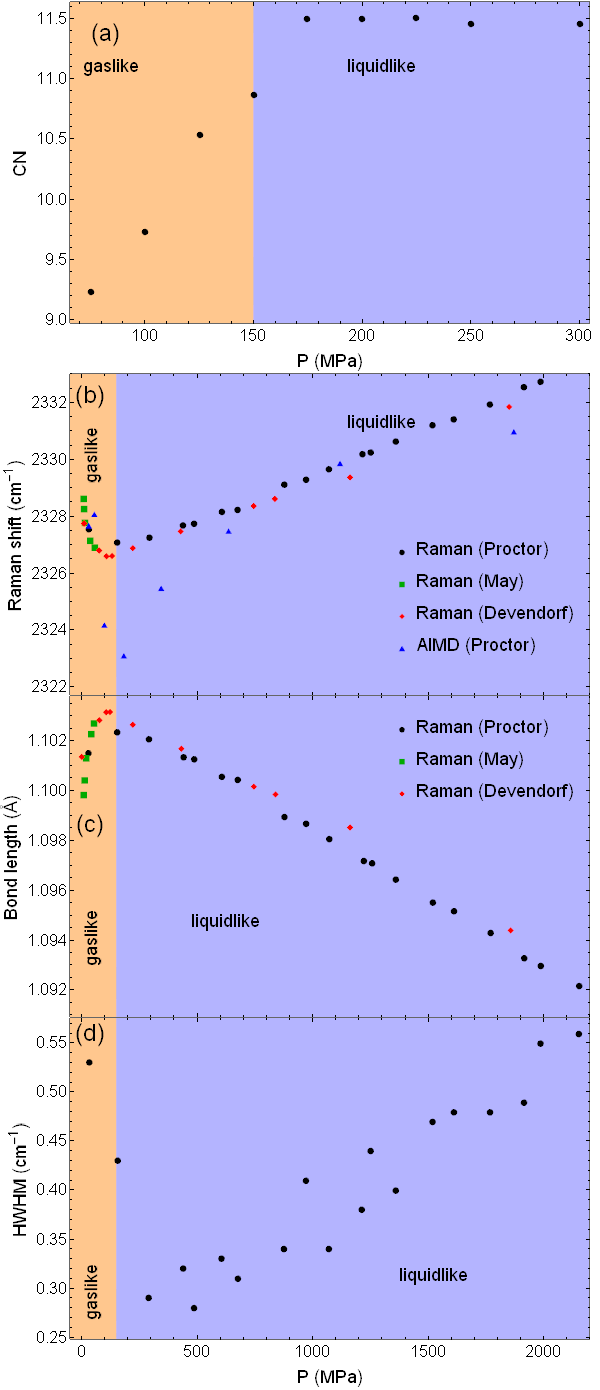}}}
\end{center}
\caption{Structural data of supercritical N$_2$ at 300 K: (a) coordination number $CN$ found by integrating $g(r)$ from the EPSR fitting as a function of $P$; (b-d) Raman shift, N-N bond length, and Raman peak half-width at half maximum supported by ab-initio molecular dynamics and similar experiment from Refs \cite{Devendorf1993,May1970} and references therein. The shaded ``gaslike" and ``liquidlike" regions meet at the Frenkel line. The data are from Ref. \cite{Proctor2019}.}
\label{fig:n2plots}
\end{figure}

This work combined neutron scattering (NS) experiments and Raman spectroscopy (RS) experiments on supercritical N$_2$ at 300 K, the former from 25 MPa to 300 MPa, and the latter from 30 MPa to 2300 MPa. RS data were obtained and are accompanied by ab-initio molecular dynamics (AIMD). The critical point of N$_2$ is 126 K and 3.4 MPa, so we are again probing the deep supercritical state, far beyond any near-critical anomalies. This isotherm crosses the FL at 150 MPa from the thermodynamic criterion ($c_V=3k_{\rm B}$ for a diatomic molecule) \cite{Proctor2019}.

Empirical potential structure refinement (EPSR) was performed on the NS data to generate structures which best fit the $S(k)$ data. The PDFs $g(r)$ were extracted directly from these simulated structures and were used to compute the mean coordination number of the sample as per equation \ref{eqn:cndef}. This is displayed in Fig. \ref{fig:n2plots}, where we observe similarity to Fig. \ref{fig:neonplots}a - molecular coordination responds much less readily to changes in pressure in the closely-packed state in the liquidlike region beyond the FL. The transition here is sharper here than in neon. This similarity is not unexpected - dinitrogen's triple bond makes it small and inert which prohibits substantial deviation from a noble fluid in terms of intermolecular structure.

The properties of N-N bond are probed by RS experiments and AIMD simulations in Ref \cite{Proctor2019} and from RS experiments in Refs \cite{Devendorf1993,May1970}. We first plot the Raman shift in Fig. \ref{fig:n2plots}b, where we observe a clear minimum at the FL, corresponding to a minimum in the frequency peak position in the vibration spectrum. Likewise the N-N bond length is maximised at the FL, though like with the Raman shift, the changes are small relative to the absolute value of the quantity in question. The breadth of the vibration mode, quantified by the half-width and half maximum (HWHM) of the Raman peak, is much more mutable under pressure changes, seen in Fig. \ref{fig:n2plots}d, but nonetheless experiences a minimum near the FL at 300 MPa.

Previous RS experiments \cite{Devendorf1993,May1970,Wang1973,Kroon1989,Scheerboom1996} have crossed the FL, reporting consistent results to these of Proctor \textit{et. al.} in Ref. \cite{Proctor2019}, however the theory of the FL, developed independently of these observations, sheds new light on these observations and brings them into a cohesive context. These intramolecular dynamical crossovers are sharp, and can be related to the intermolecular structural crossovers seen before. The dynamical quantities in the gaslike phase in Figs. \ref{fig:n2plots}b-c are much more sensitive to pressure than in the liquidlike phase, as we saw before. The explanation given by Proctor \textit{et. al.} is as follows. In the gaslike state, increasing pressure causes pronounced changes in density and coordination rather than compressing the molecules themselves. The increase in density strengthens van der Waals attractive forces between molecules, loosening the bond and causing the increase in bond length and decrease in vibration frequency with pressure on approach to the FL. In the rigid liquidlike state, the molecules start vibrating in a solid-like structure and are unable to rearrange themselves during time $\tau$. In this state, the molecules themselves are compressed as density increases, and the bond is stiffened, leading to the decrease in bond length and increase in frequency with pressure. Therefore, the minimum is expected at the crossover between gaslike and liquidlike regimes, which is the crossover at the Frenkel line.

We observe that N$_2$ is an interesting system because it's a small, weakly interacting molecule which nonetheless possesses internal degrees of freedom. It therefore serves as a good demonstrative system, much like neon did, to characterise the intramolecular transitions which the FL implicates. The overall conclusion of Ref. \cite{Proctor2019} is that dynamical and structural crossovers in intra and intermolecular properties coincide at the FL.

The following work of Pruteanu \textit{et al} \cite{Pruteanu2021} has studied the supercritical N$_2$ using neutron scattering at lower temperature of 160 K. The predicted transition at the Frenkel line was found at 85 MPa, evidenced by the crossover of coordination numbers similar to that shown in Fig. \ref{fig:n2plots}a at 300 K.

\subsection{Methane}

Our next case study is supercritical methane, CH$_4$, whose critical point is 191 K and 4.6 MPa. Unlike the experiments performed on neon and dinitrogen, here the authors of Ref. \cite{Smith2017} perform a suite of RS experiments along four different isotherms which cross the FL (calculated from the VAF criterion in earlier work \cite{Yang2015}), corresponding to the state points (298 K, 16 MPa), (345 K, 22 MPa) , (374 K, 24 MPa), (397 K, 28 MPa). These state points therefore correspond to 1.5-2$T_c$ and 35-60$P_c$.

\begin{figure}[H]
\begin{center}
{\scalebox{0.39}{\includegraphics{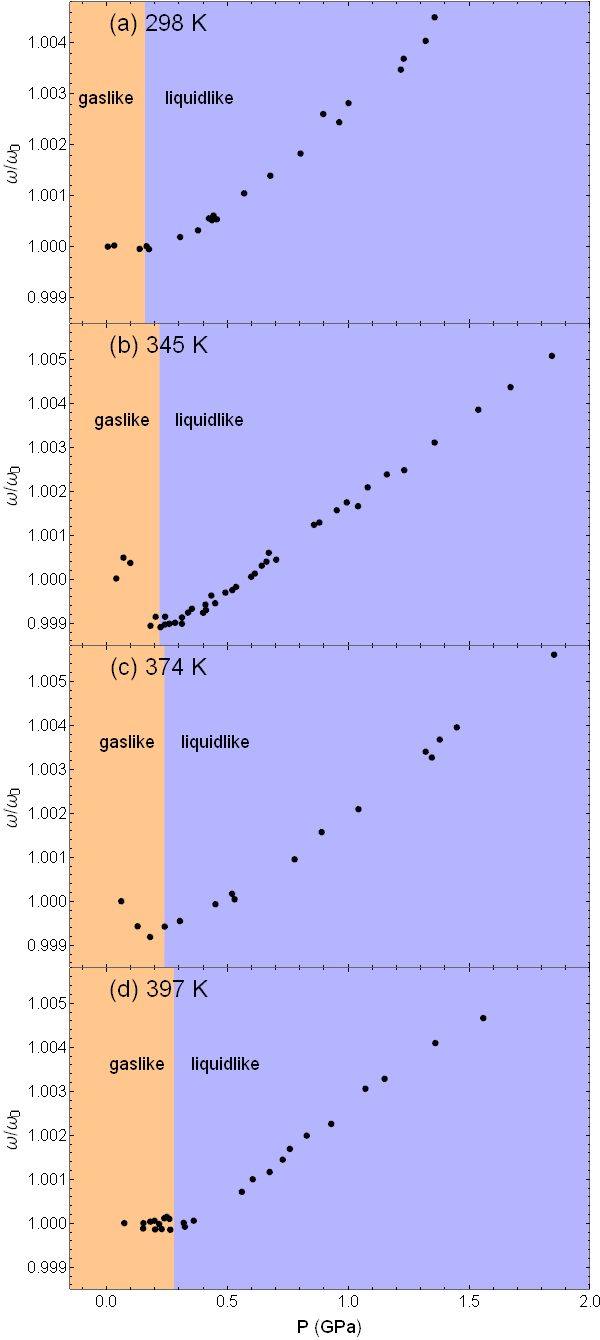}}}
\end{center}
\caption{Reduced Raman frequency shift of supercritical CH$_4$ at (a) 298 K; (b) 345 K; (c) 374 K; (d) 397 K. The shaded ``gaslike" and ``liquidlike" regions meet at the FL. The frequency $\omega_0$ corresponds to the frequency of the lowest pressure state point for each isotherm. The data are from Ref. \cite{Smith2017}.}
\label{fig:ch4plots}
\end{figure}

Methane is a tetrahedral molecule, the simplest hydrocarbon with a C-H bond length of about 1 \AA. Though this is compact, methane is larger than N$_2$ and more importantly has much richer internal dynamics (9 modes in total) due to its four C-H bonds. However of the Raman active modes, only one is intense enough for reliable data collection - the symmetric stretching vibration mode.

The reduced frequency of the symmetric stretching vibration mode is plotted as a function of pressure on all four isotherms in Fig. \ref{fig:ch4plots}. As with N$_2$ discussed in the previous section, we clearly see the minima of Raman frequencies at the FL. These striking plots manifest the discussion of the relation between intermolecular and intramolecular dynamics discussed above - in the gaslike state, attractive forces soften the C-H bond, in the closely packed liquidlike state the repulsive forces stiffen it.

Supercritical fluids, and methane in particular, are of special interest due to their existence in the atmospheres of gas giants such as Uranus and Neptune. On Uranus, methane is the third most abundant substance and can be found in its supercritical state in the lower atmosphere \cite{Helled2020a}. Anomalous heat measurements \cite{Pearl1990} and storm activity \cite{Pater2014} have been attributed to a layer of low conductivity in the atmosphere \cite{Guillot1995}. Thermal conductivity is among the many fluid properties which changes its behaviour across the FL \cite{Brazhkin2012a}, therefore comprehensive investigation into the FL of methane and other fluids present on Uranus such as water, ammonia, and hydrogen is of critical importance to planetary science. The crossover of these and other transport properties will be discussed in Section IIIB in detail.

\subsection{Ethane}

Ethane, C$_2$H$_6$, is our next case study, and is important because this experiment in Ref. \cite{Proctor2018} was conducted at a \textit{subcritical temperature}, contrary to the deeply supercritical conditions of the previous experiments. Recall that the FL is independent of the critical point, and exists in systems without a critical point or a boiling line. We also recall that in systems with the boiling line and critical point, the FL continues below the critical point and touches the boiling line below the critical temperature \cite{Brazhkin2013}. Crossing the FL by increasing the temperature at a fixed pressure below the critical point still corresponds to the transition between gas-like and liquid-like dynamics; crossing the boiling line on further temperature increase corresponds to a phase transition between low- and high-density gases because a cohesive liquid state ceases to exist close to the critical point \cite{Stishov1993}.

The experiment in ethane corroborates the FL as an entity which, though it extends into and provides structure to the supercritical state, is more fundamental than the critical point. Additionally, ethane is a molecule more complex than those in previous cases, composed of two CH$_3$ sites connected by a C-C bond. Its structure is much further from spherical than methane and possesses far richer dynamics, both of which add many more layers of complexity to the intermolecular translational motion which the FL implicates.

RS experiments were performed on ethane at 300 K from 4.4 MPa to the melting pressure of 2500 MPa \cite{Proctor2018}. The critical point of ethane is 305 K, 4.87 MPa, demonstrating that any crossovers along this isotherm certainly cannot converge onto the critical point from above. Ethane, possessing 8 atoms, expresses $3N-3 = 21$ internal modes. This is very many, and not all are equally sensitive to the FL. We select a subset which undergoes crossovers here for brevity and demonstration, but there are additional crossovers in Ref. \cite{Proctor2018}.

\begin{figure}
\begin{center}
{\scalebox{0.4}{\includegraphics{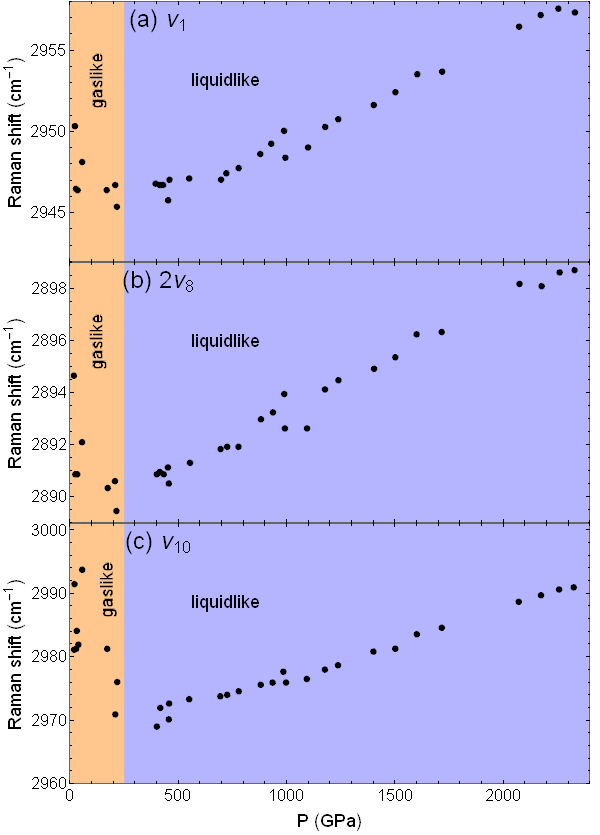}}}
\end{center}
\caption{Raman shift of supercritical C$_2$H$_6$ at 300 K of the (a) $\nu_1$; (b) 2$\nu_8$; (c) $\nu_{10}$ modes. The shaded ``gaslike" and ``liquidlike" regions meet at the approximate. The data are from Ref. \cite{Proctor2018}.}
\label{fig:c2h6flplots}
\end{figure}

We begin with the $\nu_1$, $2\nu_8$ and $\nu_{10}$ modes, which represent symmetric stretching, deformation and degenerate stretching respectively of the CH$_3$ groups (see the supplementary material of Ref \cite{Read2020} for a visualisation of these different modes), and plot the Raman shift associated with these modes along the isotherms in Fig. \ref{fig:c2h6flplots}. The minima resemble those of previous systems, occurring at 250 MPa, but we note that in this case there is no prediction of the FL from MD or other sources. The authors observed the crossover in the vicinity of 250 MPa and associated it with the FL on the basis of the previous experiments showing similar results \cite{Prescher2017,Smith2017}. This provides evidence of gaslike states, like those seen in N$_2$ and CH$_4$, at higher pressures than the boiling line but at a \textit{lower} temperature than the critical point.

\begin{figure}
\begin{center}
{\scalebox{0.4}{\includegraphics{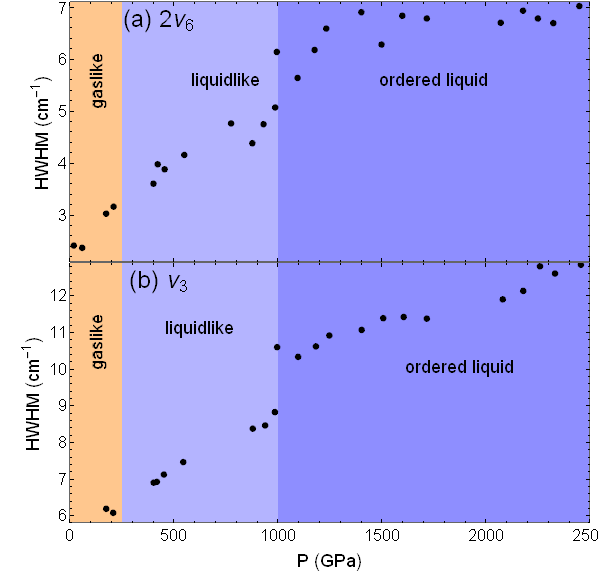}}}
\end{center}
\caption{Reduced Raman frequency shift of supercritical C$_2$H$_6$ at 300 K of the (a) 2$\nu_6$; (b) $\nu_3$ modes. The shaded ``gaslike" and ``liquidlike" regions meet at the FL. The frequency $\omega_0$ corresponds to the frequency of the lowest pressure state point for each isotherm. The data are from Ref. \cite{Proctor2018}.}
\label{fig:c2h6flplotsb}
\end{figure}

Fig. \ref{fig:c2h6flplotsb} shows the peak width at half maximum of the $2\nu_6$ and $\nu_3$ modes, representing symmetric CH$_3$ deformation and C-C bond stretching respectively. The authors of Ref. \cite{Proctor2018} note that these spectra show crossovers at 1000 MPa, deep in the liquidlike state. We have labelled the region beyond this transition an ``ordered liquid", as it lies before the melting line and is brought about by increased pressurisation and causes changes to the C-C bond. Indeed, the authors in Ref. \cite{Proctor2018} calculate the maximum density that rigid ethane can attain without any orientation order, and find that this density is exceeded beyond this liquid-liquid transition. We show it here because this transition is only possible once the ethane had transitioned to the liquidlike state from the gaslike state at lower pressures - the liquidlike state beyond the FL sports short-range order, giving it the potential to support further transitions.

\subsection{Carbon dioxide}

We now come to CO$_2$, which is one of the most commonly used supercritical fluids and is the main component (97\%) in the atmosphere of Venus. Supercritical CO$_2$ is used in a wide variety of applications such as polymer synthesis and processing \cite{Sarbu2000,DeSimone1992,Cooper2000}, dissolving and deposition in microdevices \cite{Blackburn2001}, solvation, green chemistry, green nanosynthesis and green catalysis \cite{DeSimone2002,Eckert1996,Li2008,Leitner2002,Anastas2002,Anastas2010,Beckman2004,Cole-Hamilton2003,Dahl2007,Jessop1996,Jessop1999}, extraction \cite{Reverchon1997}, chemical reactions \cite{Savage1995}, and sustainable development including carbon capture and storage \cite{Song2006}.

\begin{figure}
\begin{center}
{\scalebox{0.4}{\includegraphics{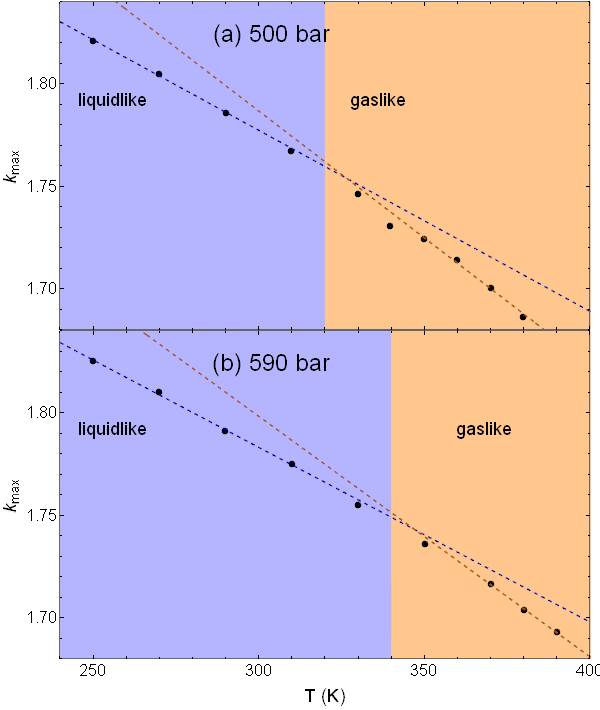}}}
\end{center}
\caption{First peak position $k_{\rm{max}}$ of structure factor $S(k)$ of supercritical CO$_2$ as a function of temperature $T$ at (a) 500 bar; (b) 590 bar. The data are from Ref. \cite{Cockrell2020a}.}
\label{fig:co2sq}
\end{figure}

Unlike previous experiments, the experimental work \cite{Cockrell2020a} studied CO$_2$ along two isobars, 500 bar and 590 bar. The FL, determined from the VAF criterion in MD simulations, passes through these isobars at 297 K and 302 K, respectively. The critical point of CO$_2$ is 304 K, 73.8 bar, again giving us evidence of the FL near the critical temperature but far exceeding the critical pressure, demonstrating that the FL is an entity independent of the critical point. NS experiments were performed on the CO$_2$ sample and supporting classical MD simulations were performed at the same conditions.

We begin with the total structure factor, $F(k)$, obtained from the NS experiments \cite{Cockrell2020a}, plotting the $k$-space position $k_{\mathrm{max}}$ of the first peak of $F(k)$ in Fig. \ref{fig:co2sq}. Similar to the results in neon in Fig. \ref{fig:neonplots}a, we see a change in the slope of $k_{\mathrm{max}}$ with temperature, occurring at temperatures 7\% and 12\% higher than the FL predicted by MD at 500 bar and 590 bar respectively. We note that the transition at the FL is broad, and the VAF criterion is only approximate and depends on the accuracy of the MD potential used, therefore these discrepancies are not unexpected. The transition remains at near-critical temperatures and very supercritical pressures, far from the Widom line at the same temperature, so that the above remarks remain appropriate.

\begin{figure}
\begin{center}
{\scalebox{0.37}{\includegraphics{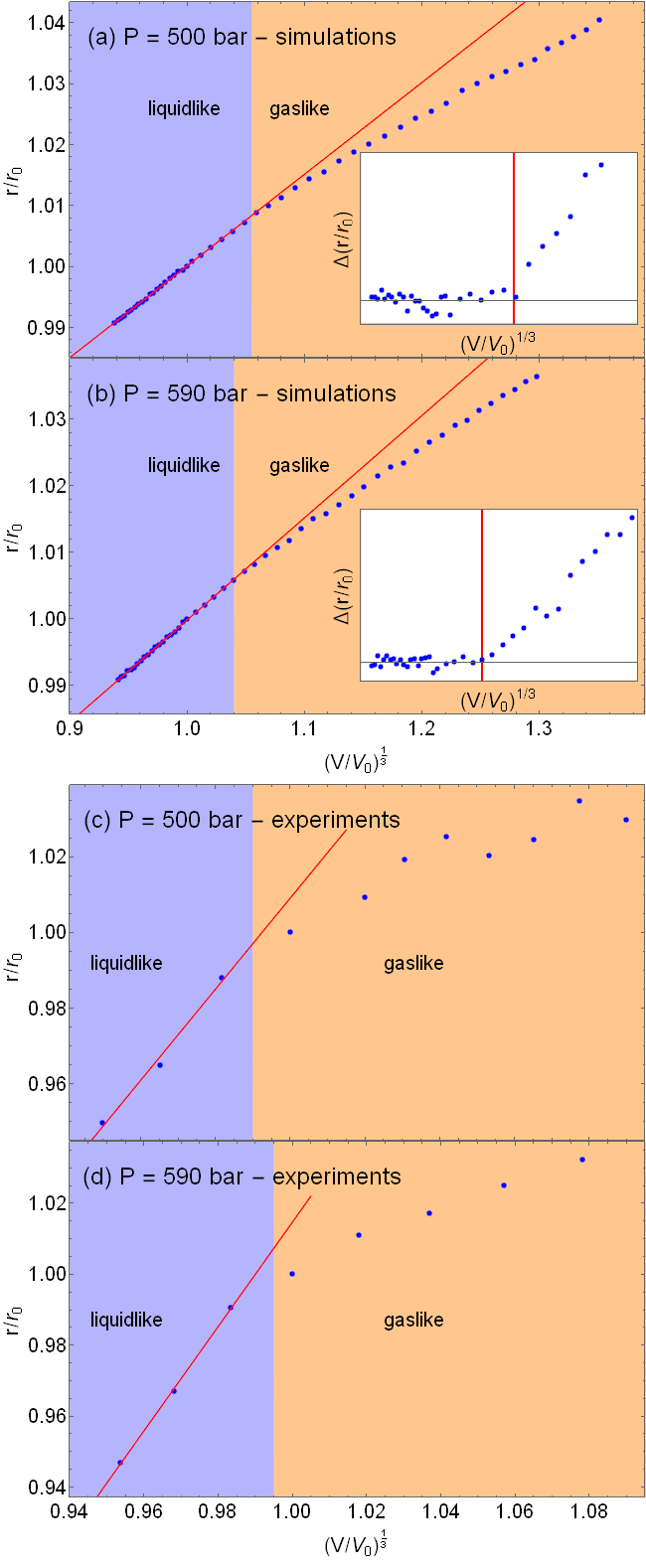}}}
\end{center}
\caption{First peak position of simulated C-C PDF of supercritical CO$_2$ at (a) 500 bar (b) 590 bar as a function of volume. The straight lines are fitted to data below the FL and serve as visual guides. The insets show the relative trend of the residuals of the linear fit; First C-C peak positions of experimental total PDF of CO$_2$ at (c) 500 bar (d) 590 bar as a function of volume. The straight lines are visual guides. The data are from Ref. \cite{Cockrell2020a}.}
\label{fig:co2peakpos}
\end{figure}

The natural structural data extracted from MD is the PDF, which we now turn our attention to. The C-C partial PDF gives the mean relative distance between molecules (between centers of mass), and its first peak is therefore the first nearest-neighbour (fnn) distance, $r_{\rm{fnn}}$, between molecules. If a system's structure doesn't change when it expands or contracts, $r_{\rm{fnn}}$ will be proportional to the system's ``length", $V^{1/3}$. This is called uniform compression, because the nearest-neighbour distance changes only due to changes in density. The proportionality constant will depend on the system's structure and on the existence of more condensed phases at higher densities (see Ref. \cite{Cockrell2020a} for a detailed discussion). If, on the other hand, the system's structure changes due to coordination rearrangements or a change in orientation order, for example, $r_{\rm{fnn}}$ will experience changes from more sources than just the change in density, and will therefore not be proportional to $V^{1/3}$.

The dependence of the ratio $r_{\mathrm{fnn}}/r_0$ (henceforth abbreviated to $r$) on $(V/V_0)^{1/3}$, where $r_0$ and $V_0$ are the first nearest-neighbour distance and volume respectively at a reference state point, from MD and NS experiments are plotted in Fig. \ref{fig:co2peakpos}. We see at higher densities that $r/r_0$ is proportional to $(V/V_0)^{1/3}$, implying uniform compression. This linear relationship has been experimentally
observed in molten group 1 elements \cite{Tsujia1996, Katayama2003} and liquid CS$_2$ \cite{Yamamoto2006}. At lower densities, this linear relationship is broken, representing a change in structure, consistent with a gaslike phase where the nearest neighbour distance is due to transient clustering rather than a solid-like condensed bulk which is directly sensitive to the bulk density. Figs. \ref{fig:co2peakpos}c-d are calculated from the total pair distribution function, the Fourier transform of $F(k)$ from the NS data. However the contribution from the first C-C peak is not significantly changed and therefore gives a qualitative approximation to $r$. The MD data have much reduced noise, and sample many more state points, making the crossover much clearer and sharper in these data. This allowed us in Ref. \cite{Cockrell2020a} to perform statistical analysis, which decisively confirmed a crossover in the functional dependence of $r/r_0$ on $(V/V_0)^{1/3}$. In this sense the NS data corroborate the structural crossover in real space, rather than unequivocally demonstrating it alone. In all four cases the crossover corresponds to a temperature within 15\% of the predicted FL from the VAF criterion.

Another set of experiments employed small-angle neutron scattering (SANS) experiments to study droplet formation in supercritical CO$_2$. It was found that this process operates at a borderline of thermodynamic parameters identified with the Frenkel line \cite{Pipich2018,Pipich2020}.

\subsection{Water}

We now come to our final case study, supercritical water. We note that experiments in supercritical Ne, N$_2$, CH$_4$, C$_2$H$_6$ and CO$_2$ discussed above were stimulated by the FL idea and were guided by the preceding calculations of the FL in these systems. Water is an exception in the sense that we analyse the experimental data collected before the FL was proposed.

Due to its prominence and significance, water research is a separate area on its own. More recently, research into the supercritical water has been on the rise due to its industrial applications, including in dissolving, cleaning and extracting processes \cite{Kiran2000a,McHardy1998}. Despite its importance, water is probably not an ideal system to study theoretical predictions: it is a complicated structure with many anomalies \cite{Gallo2016}, one of which is the gradual increase of coordination numbers with temperature.

Compared to subcritical water, relatively few experiments were done at high pressure and high temperature. For example, structure of water was studied in X-ray diffraction experiments up to 17 GPa and 850 K \cite{Ikeda2010}. The state points followed the melting line lying below the FL. Soper's pioneering neutron scattering experiments in supercritical water included the range of the FL \cite{Soper2000}, although the state points were fairly sparse for an analysis of a transition across the FL.

An earlier X-ray study \cite{Gorbaty1983} explored 11 temperature points in supercritical water at 1 kbar. Note that the Widom line no longer exists at these kbar pressures. The state points in Ref. \cite{Gorbaty1983} were inside a pressure window where a structural transformation of water operates and where the effect of the FL was predicted to be most noticeable \cite{Cockrell2020}: classical molecular dynamics simulations indicated that the FL crossover in supercritical water is close to the structural crossover between differently coordination states. For this reason, we compare these experiments to the predictions of the transition at the FL.

\begin{figure}
\begin{center}
{\scalebox{0.35}{\includegraphics{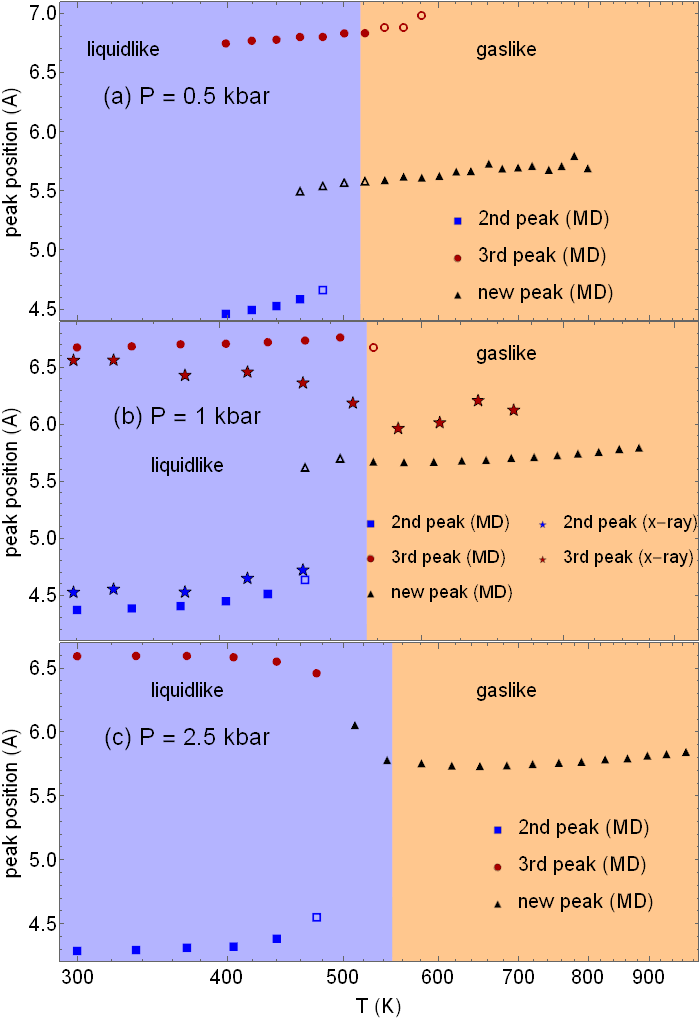}}}
\end{center}
\caption{O-O PDF peak positions from molecular dynamics (MD) data \cite{Cockrell2020} and X-ray scattering data \cite{Gorbaty1983} at (a) 0.5 kbar; (b) 1 kbar; (c) 2.5 kbar. Open triangles imply that the new peak in the MD data is less prominent than its neighbours and vice versa for open squares and circles.}
\label{fig:waterpeakpos}
\end{figure}

In Fig. \ref{fig:waterpeakpos} we plot the positions of the secondary peaks of the O-O PDFs from MD at three pressures. The temperatures of the FL, determined from the VAF criterion in the previous work \cite{Yang2015}, are shown as vertical boundaries. The crossover in the MD data is pronounced at 0.5 kbar and 1 kbar - the second and third peaks diminish in height and disappear. Concomitantly, a new peak at a radial distance intermediate to the old peaks emerges and becomes more prominent. At higher 2.5 kbar pressure, the crossover is less abrupt and is seen as the second peak disappearing and the third peak evolving into the peak close to the FL.

In Fig. \ref{fig:waterpeakpos}b we also plot the experimental XS data \cite{Gorbaty1983}. We observe that the positions of the second and third peaks in MD and XS at 1 kbar are close to each other at low temperature. The experimental second peak abruptly disappears at the same temperature as the MD peak and close to the FL temperature. The third experimental peak is close to the MD peak at low temperature, starts to deviate from it at higher temperature and approaches the new MD peak at temperature close to the FL. This behavior of the third peak is close to that in the MD simulations at higher pressure in Fig. \ref{fig:waterpeakpos}c - it drops down to a minimum close to the FL and then moves outwards upon further temperature increase. The structural crossover and its proximity to the FL predicted by MD simulations, while less abrupt, is therefore present in these experimental data.

This structural crossover in water is well-known: subcritical water is known for its tetrahedral, hydrogen-bonded intermolecular structure. At higher temperatures and pressures water loses this structure and instead adopts a closely-packed structure \cite{Ikeda2010,Chara2011,Marti1999}. The disappearance of the second and third peaks and the appearance of a new intermediate peak is the footprint of this crossover. Based on these observations, it appears that the FL is a facilitator of the crossover between the tetrahedral and closely-packed structures: as the solid-like oscillatory component of molecular motion is lost in the tetrahedral structure, water molecules acquire purely diffusive gaslike motion and hence flexibility to arrange into a denser structure in response to high pressure. This near coincidence of the dynamical crossover with the structural crossover occurs in MD simulations at 6 pressures from 0.5 kbar to 10 kbar \cite{Cockrell2020}.

We note that the MD data rely on the empirical potential used. Although the potential used in Ref. \cite{Cockrell2020} is considered to be of high quality, the empirical potentials are known to work well for some system properties and describe others with less precision, especially if those are subtle. Improved and detailed MD simulations of supercritical water, perhaps involving ab initio simulations, are likely to improve the agreement between experiments, simulations and theory. In this context, it is interesting to note recent modelling work \cite{Skarmoutsos2021} reporting a structural crossover in the supercritical water at the Frenkel line using a different interatomic potential. In addition to the structural crossover, this work detected a crossover of heat capacity around 1 GPa which, according to the earlier prediction in Ref. \cite{Yang2015}, corresponds to the FL in supercritical water.

Further detailed experiments in supercritical water would certainly be important and interesting in order to further specify the nature and extent of the structural crossover in this state, its relationship with the dynamical crossover at the FL, and other related consequences of them both. These questions are important from the point of view of fundamental understanding of this system as well as its industrial applications \cite{Kiran2000a,McHardy1998}.

\subsection{Summary}

In summary, the experiments discussed in this section show a variety of ways in which the transitions in the supercritical state can be studied. They also illustrate a wide chemical and structural variety of fluids, including those with critical industrial and environmental applications, exhibiting supercritical transitions. As we survey and chart the hitherto unknown deep supercritical state, further experiments will continue to be important and reveal new physics in supercritical fluids. This research will benefit from recent experimental advances in scattering techniques (see, e.g. Ref. \cite{Petrillo2021}) shaping the future of supercritical research.

\section{Related, developing and future research}

\subsection{Dynamical effects}

Although this review is primarily experimental, it is nevertheless interesting to mention different theoretical and modelling approaches stimulated by the FL idea, with the view that these results may suggest new future experiments as well as stimulate further theoretical and modelling developments. We do not intend to provide an exhaustive list but rather indicate a range of areas and topics that followed since the introduction of the FL which holds promise for future progress.

From the point of view of FL analysis and mapping, we note an extended and detailed study of dynamics, transverse modes and VAFs using mode decomposition \cite{Belissima2017}. This study confirmed the basic physics involved in the dynamical crossover at the FL \cite{Belissima2017}. An interesting question arises of how can the FL idea and VAF analysis be extended in the quantum regime. A notable development is to apply the FL analysis to path integral molecular dynamics simulations to study quantum effects in supercritical helium \cite{Takemoto2018}. VAF and the crossover at the FL were calculated and discussed in that work. It was suggested that the FL universally applicable to both classical and quantum fluids so that future simulations can map the FL in other quantum fluids \cite{Takemoto2018}.

An interesting development is the application of the FL to unanticipated physical scenarios and systems. The FL idea was used to discuss the topological and geometrical framework and percolation of solid-like clusters \cite{Yoon2019,Yoon2018}. This relation to the FL was later developed to discuss the implications for the system entropy, isomorphism and scaling \cite{Yoon2019a}. Another modelling study \cite{Ghosh2019} found that different dynamical regimes below and above the FL were found to have non-trivial effects in confined supercritical fluids in terms of structure, diffusion and entropy, leading to verifiable predictions. Different properties of confinement were found to affect structure and layering in the liquidlike regime below the FL but not in the gaslike regime above the FL \cite{Ghosh2018,Ghosh2019a}. The crossover at the FL in confined supercritical fluids was also shown to be related to the changes of structure, diffusion and dynamics including the disappearance of transverse modes \cite{Ghosh2019a}.

\subsection{Transport properties and fundamental bounds}

As mentioned in section IB earlier, the crossover of particle dynamics from combined oscillatory and diffusive to purely diffusive results in different behavior of important properties of the supercritical system in these dynamically distinct states. For example, viscosity, speed of sound, and thermal conductivity all decrease with temperature in the liquidlike regime but increase with temperature above the FL as in gases \cite{nist}. Therefore, these properties all have minima. The diffusion constant crosses over from exponential temperature dependence in the liquidlike regime to power-law dependence as in gases \cite{Brazhkin2012}. Depending on the path chosen on the phase diagram, the crossovers and minima of these properties may deviate from the FL but nevertheless lie fairly close to it \cite{Brazhkin2012b,Fomin2018path}.

\begin{figure}
\begin{center}
{\scalebox{0.45}{\includegraphics{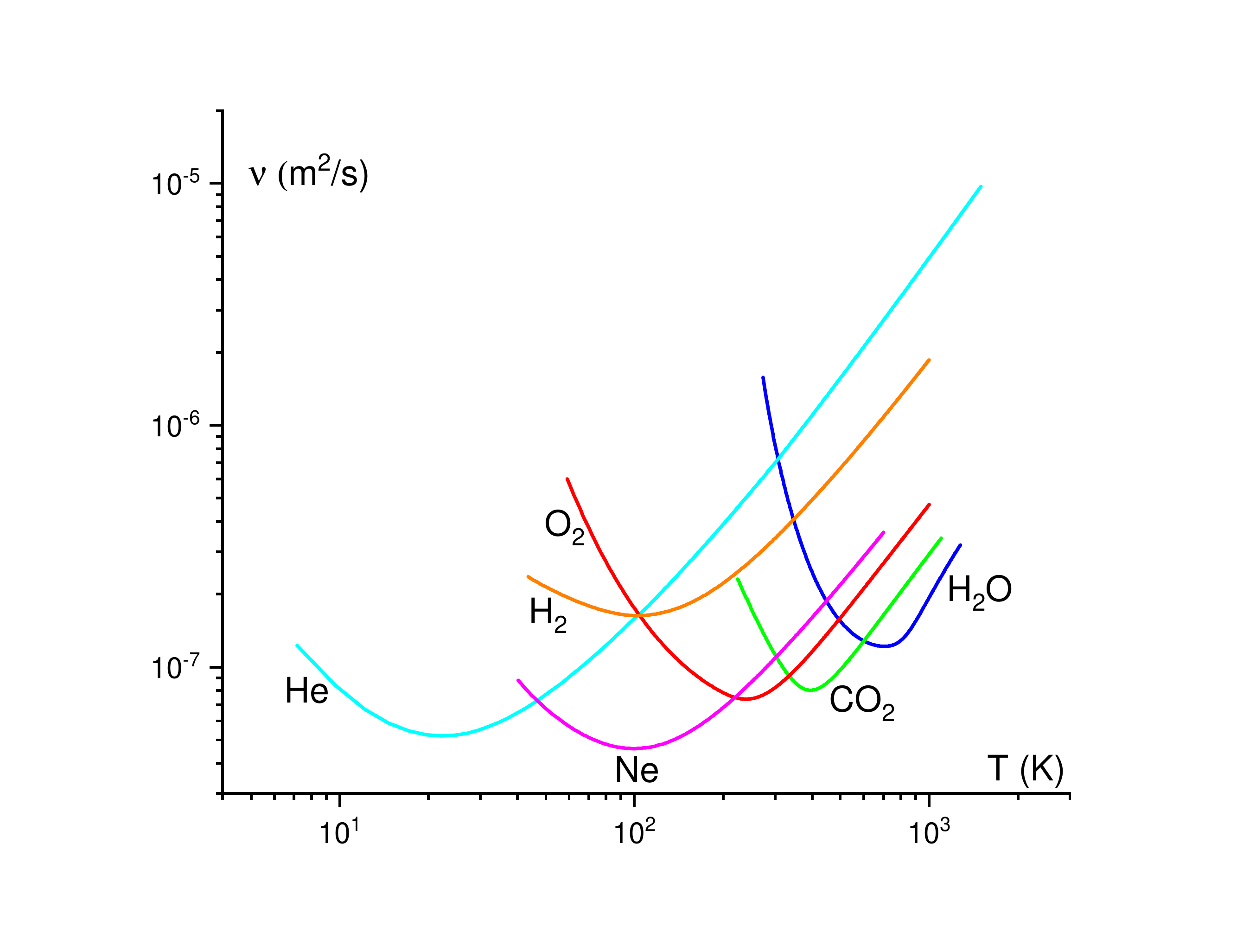}}}
\end{center}
\caption{Experimental kinematic viscosity of supercritical molecular, noble and network fluids showing minima \cite{nist}. $\nu$ for H$_2$, He, O$_2$, Ne, CO$_2$ and H$_2$O are shown at 50 MPa, 20 MPa, 30 MPa, 50 MPa, 30 MPa and 100 MPa, respectively.}
\label{nu}
\end{figure}

As an important example of the supercritical crossover, we show the experimental kinematic viscosity \cite{nist} $\nu=\frac{\eta}{\rho}$ ($\eta$ is viscosity and $\rho$ is density) in Fig. \ref{nu} and observe the minima just mentioned.

\begin{figure}
\begin{center}
{\scalebox{0.45}{\includegraphics{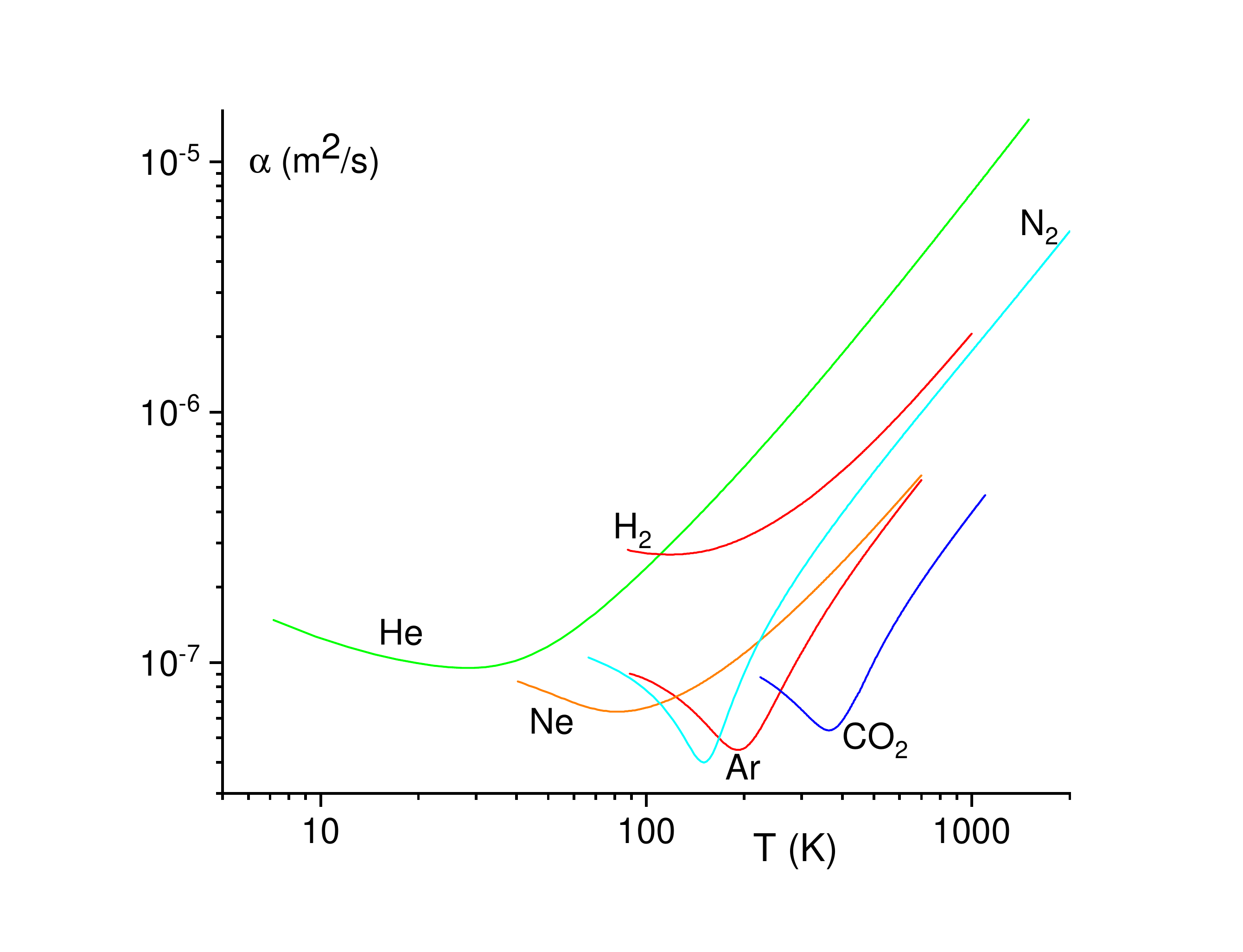}}}
\end{center}
\caption{Experimental thermal diffusivity $\alpha$ of supercritical noble and molecular fluids \cite{nist} showing minima. $\alpha$ for Ar, Ne, He, H$_2$, N$_2$ and CO$_2$ are shown at 20 MPa, 50 MPa, 20 MPa, 100 MPa, 10 MPa and 30 MPa, respectively.}
\label{alpha}
\end{figure}

Minima are also seen in another important property responsible for heat transfer, thermal conductivity and thermal diffusivity $\alpha=\frac{\kappa}{\rho c_p}$ ($\kappa$ is thermal conductivity and $c_p$ is isobaric specific heat capacity). We show the experimental $\alpha$ \cite{nist} in Fig. \ref{alpha} and, similarly Fig. \ref{nu}, observe the minima.

Interestingly, the values of $\nu$ and $\alpha$ at their minima, $\nu_m$ and $\alpha_m$, turn out to be nearly universal and largely governed by fundamental physical constants \cite{Trachenko2020,Trachenko2021}. It has been shown that $\nu_m$ and $\alpha_m$, related to the dynamical crossover, are governed by interatomic spacing and Debye frequency. Expressing these in terms of fundamental physical constants gives

\begin{equation}
\nu_m=\alpha_m=\frac{1}{4\pi}\frac{\hbar}{\sqrt{m_em}}
\label{minima}
\end{equation}

\noindent where $m_e$ and $m$ are electron and molecule masses, in agreement with a wide range of experimental data in supercritical fluids \cite{Trachenko2020,Trachenko2021}.

Eq. \eqref{minima} answers the long-standing Purcell question \cite{Purcell1977}, namely why liquid viscosity never falls below a certain value? In the first footnote of Ref. \cite{Purcell1977}, Purcell mentioned that Weisskopf has explained this to him. Although we did not find a published explanation, it is worth noting that the same year Weisskopf published the paper ``About liquids'' \cite{weisskopf}. That paper starts with a story often recited by conference speakers: imagine a group of isolated theoretical physicists trying to deduce the states of matter using quantum mechanics only. They are able to predict the existence of gases and solids, but not liquids. Weisskopf implies that liquids are hard to understand. We will explain the origin of this difficulty in the next section IIIC. Here, we note that this problem applies to dense and strongly-interacting supercritical fluids to the same extent as it does to subcritical liquids.

The Purcell question can be answered with the aid of Eq. \eqref{minima} as follows. Liquid viscosity never falls below certain value because (a) fluids universally have a minimum related to the crossover from the liquidlike to gaslike dynamics as seen in Fig. \ref{nu} and (b) the minimum for each liquid is fixed by the fundamental physical constants in Eq. \ref{minima} and does not vary too much across liquids due to a fairly weak $\propto\frac{1}{\sqrt{m}}$ dependence.

More generally, it should be noted that finding and understanding bounds on physical properties, including transport properties, has enthralled physicists for quite a long time and improved our understanding of classical and quantum dynamics including collective properties (see Ref. \cite{Grozdanov2021} for a short recent review). One notable example is the Kovtun-Son-Starinets (KSS) lower bound on the ratio of viscosity and entropy derived using string theory methods and involving black hole physics \cite{kss}. This involved the holographic correspondence between strongly-interacting field theory and its gravitational dual and, interestingly, compared the derived bound to the minima in fluids such as those shown in Fig. \ref{nu}. The comparison revealed the KSS bound to be much lower that the minima in fluids. This difference was later attributed to the proton-to-electron mass ratio \cite{Trachenko2020}. We note that bounds on different transport properties are often heuristic and involve uncertainty relations where a function in question does not necessarily have a minimum \cite{Grozdanov2021,kss}. In interesting difference to this approach, the lower bounds on viscosity and thermal conductivity (diffusivity) in supercritical fluids in Figures \ref{nu} and \ref{alpha} represent the real minima of corresponding functions related to the dynamical crossover.

We note that Fig. \ref{nu} and \ref{alpha} include supercritical H$_2$ and He important for the physics of gas giants such as Jupiter and Saturn. We will discuss the implications of the FL in hydrogen for astrophysics and planetary science in chapter IIID below.

\subsection{Thermodynamic theory}

Using recent results and in particular experimental data discussed above, it is important to discuss theoretical understanding of the deeply supercritical state. Existing discussions of thermodynamic properties were mostly limited to the vicinity of the critical point only \cite{Sengers1986,Anisimov2011}.

Notably, thermodynamic theory of liquids in general has remained a long-standing problem in physics due to the combination of strong interactions and large particle displacements. For this reason, it is believed that no general results for liquid thermodynamic functions (energy, heat capacity and so on) can be derived theoretically, in contrast to solids and gases \cite{Landau1980}. According to Landau and Lifshitz, the liquid energy is strongly system-specific because inter-particle interactions are both strong and system-dependent, and this precludes the calculation of liquid energy in general form \cite{Landau1980}. As a result, very little or nothing is said about liquid specific heat in textbooks, representing a challenge for undergraduate teaching \cite{Granato2002}.

We note the earlier Weeks-Chandler-Andersen (WCA) approach, which emphasised the role of repulsive interactions in setting the structure of certain class of liquids at high density and used this insight to expand the liquid free energy around the reference state given by repulsive forces in the perturbation theory where the attractive term is a perturbation \cite{Chandler1983,Weeks1971}. Calculating the energy numerically as an integral over interaction energy and correlation functions gives the main thermodynamic result of the WCA approach, specifically that the repulsive part of the interaction potential plays the dominant role in setting the liquid energy for certain fluid types at high density \cite{Weeks1971}. If applied to understanding liquid energy and heat capacity in general form and their temperature dependence, the WCA approach suffers from the same problem emphasised by Landau and Lifshitz: the energy integral (even if could be evaluated analytically rather than numerically) relies on the explicit knowledge of the liquid structure and interatomic interactions, the system-specific properties. As a result, the energy becomes strongly system-dependent, making the approach unsuitable for a broad understanding of liquids. The WCA approach is therefore unable to provide physical insights into most important and general thermodynamic liquid properties such as the universal decrease of liquid specific heat from its Dulong-Petit value of $3k_{\rm B}$ to the ideal-gas value of $\frac{3}{2}k_{\rm B}$ with temperature \cite{Trachenko2016}. This is not a criticism of the WCA approach because the WCA theory did not intend to explain general thermodynamic liquid properties mentioned above, but was mostly aimed at ascertaining the relative contributions of repulsive and attractive forces to liquid structure and energy for certain liquid types at different densities \cite{Chandler1983,Weeks1971}.

This same fundamental problem of theoretical description equally applies to dense and strongly-interacting supercritical fluids as it does to subcritical liquids.

More recently and with the development of new generation synchrotrons, it became apparent that liquids are able to support phonons including solidlike transverse phonons. Together with the phonon theory of liquid thermodynamics which emerged a few years earlier, the behavior of liquid specific heat was understood on the basis of propagating phonons, as is the case in solids (see Ref. \cite{Trachenko2016} for review). The important question is whether the supercritical state can be understood using the same approach.

The question of thermodynamic theory of the supercritical state has two parts to it:\\
(a) thermodynamic properties of supercritical matter below and above the FL, and\\
(b) properties at the FL itself.

We have a fairly good understanding of (a) but not of (b). For the benefit of the reader, we briefly review the discussion of supercritical thermodynamics. This will help us formulate open questions and list problems for the future.

The phonon theory of liquid thermodynamics is based on liquid's ability to support collective modes, phonons. The starting point is the Maxwell-Frenkel theory of liquids (see, e.g. Ref. \cite{Trachenko2016} for review). Maxwell proposed that the liquid response to an external perturbation (for example, shear stress) is neither hydrodynamic nor elastic, but has contributions from both effects \cite{Maxwell1867}. Frenkel formalised this by writing \cite{Frenkel1955}:

\begin{equation}
\frac{dv}{dy}=\frac{P}{\eta}+\frac{1}{G}\frac{dP}{dt}
\label{a1}
\end{equation}

\noindent where $v$ is the velocity perpendicular to $y$-direction, $\eta$ is viscosity, $G$ is shear modulus and $P$ is shear stress.

According to Eq. (\ref{a1}), the shear deformation in a liquid is the sum of the viscous and elastic deformation rates given by the first and second right-hand side terms, respectively. This has given rise to term ``viscoelasticity'' of liquids, although ``elastoviscosity'' would be an equally legitimate term because both deformations are treated in (\ref{a1}) on equal footing. Using \eqref{a1}, Frenkel modified the Navier-Stokes equation to include the elastic response. A simplified form of the resulting equation reads \cite{Trachenko2016}:

\begin{equation}
c^2\frac{\partial^2v}{\partial x^2}=\frac{\partial^2v}{\partial t^2}+\frac{1}{\tau}\frac{\partial v}{\partial t}
\label{gener3}
\end{equation}

\noindent where $v$ is the velocity component perpendicular to $x$, $c$ is transverse wave velocity $c=\sqrt\frac{G}{\rho}$, $\rho$ is density and $\tau=\frac{\eta}{G}$ is the liquid relaxation time. Frenkel associated this $\tau$ with the time between molecular jumps discussed earlier \cite{Frenkel1955}. This association has become an accepted view \cite{Dyre} and is supported by modelling results (see, e.g., Ref. \cite{Egami}).

Seeking the solution of (\ref{gener3}) as $v=v_0\exp\left(i(kx-\omega t)\right)$ gives

\begin{equation}
\begin{aligned}
&\omega=-\frac{i}{2\tau}\pm\omega_r\\
&\omega_r=\sqrt{c^2k^2-\frac{1}{4\tau^2}}
\label{kgap}
\end{aligned}
\end{equation}

In the case that $k<k_g$, where

\begin{equation}
k_g=\frac{1}{2c\tau}
\label{kgap1}
\end{equation}

\noindent is the gap in momentum space, $\omega$ in Eq. \eqref{kgap} has neither a real part nor transverse phonons. Since transverse phonons exist only for $k>k_g=\frac{1}{2c\tau}$, Eq. \eqref{kgap} results in gapped momentum states (GMS) in liquids.

GMS operate in several different areas and systems in physics, and liquids and supercritical fluids is one such example \cite{Baggioli2020}.
There is a substantial experimental evidence for propagating waves in liquids including solid-like transverse waves whose frequency and wave vectors extend to the zone boundary as in solids (see, e.g., Refs. \cite{Copley1974,Pilgrim1999,Burkel2000,Pilgrim2006,Hosokawa2009,Giordano2010,Giordano2011,Hosokawa2013,Hosokawa2015,Khusnutdinoff2020,Khusnutdinoff2020e}, however experiments have not yet demonstrated GMS in liquids in terms of gapped dispersion curves. Gapped dispersion curves were directly measured in dusty plasma only \cite{Nosenko2006}. Experimental support for GMS in liquids comes from the liquid ability to support shear stress at small length scales \cite{Noirez2012,Noirez2010}, in agreement with the predicted propagation of shear waves at large wavevectors $k>k_g$ \eqref{kgap1}. The evidence for GMS in liquids also comes from molecular simulations \cite{Yang2015,Baggioli2020}.

We now recall the first qualitative criterion of the FL discussed in section IB: $\tau\approx\tau_{\rm D}$. When $\tau$ approaches the shortest time scale in the system, $\tau_{\rm D}$, $k_g$ in Eq. \eqref{kgap1} becomes $k_g\approx\frac{1}{c\tau_{\rm D}}=\frac{1}{a}$, where $a$ is interatomic separation. This gives $k_g\approx k_{\rm D}$, or shortest wavevector in the systems, at which point all transverse modes disappear from the system's spectrum. As discussed in Section IB, this disappearance is the hallmark of the FL. Hence, the FL serves as a boundary limiting the propagation of solid-like transverse waves. More recently, the evolution and disappearance of transverse modes was discussed in detail with an account of damping effects \cite{Kryuchkov2019}.

We note that similarly to solids, plane waves decay in liquids. The decay mechanisms in solids include anharmonicity, defects and structural disorder present in, for example, glasses. Despite this decay, high-temperature specific heat in disordered glasses is governed by phonons. In liquids, the additional decay mechanism is related to atomic jumps \cite{Trachenko2016}. The propagation length of high-frequency phonon excitations in liquids and supercritical fluids is on the order of nanometers, as evidenced by experiments and modelling \cite{Hosokawa2009,Hosokawa2015,Hosokawa2013,Yang2017}. This is similar to room-temperature crystalline metals where the lifetime of high-frequency phonons is on the order of picoseconds and the propagation range is on the order of nanometers \cite{Jain2016} and where disorder and/or defects can reduce these values further. This is also similar to glasses which are structurally similar to liquids \cite{ruffle}. Therefore, phonon excitations govern the specific heat in liquids and supercritical fluids to the same extent they do in solids.

The liquid energy below the FL can now be derived by calculating the energy of two transverse modes above $k_g$ or, equivalently, the energy of propagating modes above frequency $\frac{1}{\tau}$ \cite{Yang2017,Fomin2018} (propagating modes correspond to $\omega_r\Gamma>1$ where $\Gamma=\frac{1}{2\tau}$ is decay rate from Eq. (\ref{kgap}) which gives $k>k_g\sqrt{2}$, or approximately $\omega>\frac{1}{\tau}$ from Eqs. \eqref{kgap}- \eqref{kgap1} \cite{Baggioli2020}). Adding the energy of the longitudinal phonon and the kinetic energy of diffusing atoms gives the liquid energy below the FL in harmonic classical case \cite{Trachenko2008,Trachenko2016,Yang2017,Fomin2018} as:

\begin{equation}
E=Nk_{\rm B}T\left(3-\left(\frac{\omega_{\rm F}}{\omega_{\rm D}}\right)^3\right)
\label{harmo}
\end{equation}

\noindent where $N$ is the number of particles and $\omega_{\rm F}=\frac{1}{\tau}$.

Eq. \eqref{harmo} predicts that the heat capacity decreases with temperature because $\omega_{\rm F}$ increases. As follows from the preceding discussion, this is due to progressive disappearance of transverse modes from the system's spectrum with temperature.

A rigorous and detailed test of this theory, and in particular the quantum extension of Eq. \eqref{harmo} \cite{Trachenko2016}, was independently performed in Ref. \cite{Proctor2020} where it was concluded that the theory, being falsifiable, quantitatively agrees with the experimental data of liquid specific heat for several different fluids over a wide temperature and pressure range.

Above the FL where the two transverse modes disappear, the system energy is governed by the evolution of the remaining longitudinal phonon with the wavelength larger than then particle mean-free path $L$. Adding to this energy the kinetic energy of diffusing atoms gives the energy above the FL in the harmonic classical case as \cite{Trachenko2016}:

\begin{equation}
E=\frac{3}{2}Nk_{\rm B}T+\frac{1}{2}Nk_{\rm B}T\left(\frac{a}{L}\right)^3
\label{supe1}
\end{equation}

\noindent where $a$ is the interatomic spacing. Both Eq. \eqref{harmo} and Eq. \eqref{supe1} can be adjusted to accommodate the effects of anharmonicity and thermal expansion that can be substantial in high-temperature fluids \cite{Trachenko2016}.

$L$ increases with temperature and, as a result, the specific heat decreases with temperature and tends to its ideal-gas value $c_V=\frac{3}{2}k_{\rm B}$, in agreement with experiments in high-temperature supercritical fluids \cite{Trachenko2016,nist}.

We now come to point (b) above, the thermodynamic behavior at the FL itself. Eq. \eqref{harmo} below the FL an Eq. \eqref{supe1} above the FL have different functional forms and different temperature behavior. Eq. \eqref{harmo} at $\omega=\omega_{\rm F}$ ($\tau=\tau_{\rm D}$) and Eq. \eqref{supe1} $a=L$ match at $E=2Nk_{\rm B}T$, corresponding to the specific heat

\begin{equation}
c_V=2k_{\rm B}
\label{cv2}
\end{equation}

\noindent which is the thermodynamic criterion of the FL in the harmonic classical case as discussed in section IB earlier.

The important open question is the nature of the transition between the two functions given by Eq. \eqref{harmo} and Eq. \eqref{supe1} at the crossover value \eqref{cv2}. The problem is that while temperature dependence of $\omega_{\rm F}$ in Eq. \eqref{harmo} and $L$ in Eq. \eqref{supe1} can be written for states away from the FL, they are not known analytically close the FL. Moreover, $\omega_{\rm F}=\frac{1}{\tau}$ and $L$ are parameters that are well-defined away from the FL but not close to the FL where they are defined approximately only. This represents a challenge for the theory which needs to identify a suitable order parameter and offer its analytical temperature dependence close to the FL in order to study the nature of the transition.

\begin{figure}
\begin{center}
{\scalebox{0.35}{\includegraphics{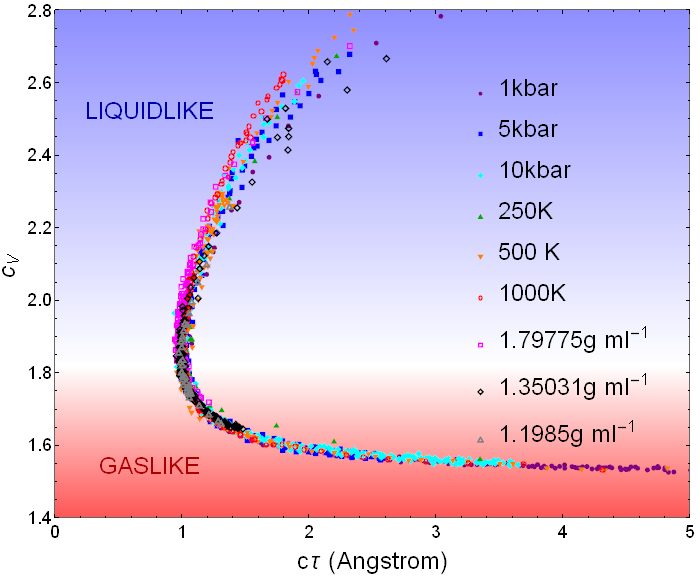}}}
\end{center}
\caption{Specific heat $c_V$ in the units of $k_{\rm B}$ as a function of $c\tau$ across 9 paths spanning the supercritical state up to 330 $T_c$ and 8000 $P_c$. $k_{\mathrm{B}}=1$. All these paths collapse onto a single curve and undergo a dynamic-thermodynamic ``c''-transition at the path-independent inversion point $c_V\approx 1.9k_{\rm B}$ and $c \tau=1~\mathrm{\AA}$. The data are from Ref. \cite{Cockrell2021}.}
\label{fig:divergence}
\end{figure}

A useful insight, and a reiteration of the special thermodynamic role played by $c_V\approx 2k_{\rm B}$ in Eq. \eqref{cv2} at the FL, comes from a recent large survey of the supercritical state extending beyond 330$T_c$ and 8000$P_c$ along isobars, isotherms, and isochores \cite{Cockrell2021}. While the representative thermodynamic property $c_V$ will always decrease towards $\frac{3}{2}k_{\rm B}$ when entering into the gaslike state, $\tau$ is more problematic, exhibiting minima or a lack of minima depending on the specific path taken when crossing between the liquidlike and gaslike phases. $c_V$ was therefore calculated as a function of the key parameter $c\tau$ entering Eq. \eqref{kgap1} along 9 phase diagram paths differing by orders of magnitude in pressure and temperature. The result is plotted in Figure \ref{fig:divergence} and shows that the curves along nine different paths exhibit the same ``c" shaped curves, or ``c''-transition, and collapse onto the exact same curve in the gaslike state below $c_V\approx 1.9k_{\rm B}$. The location of the inversion point is physically meaningful: it lies close to the predicted crossover $c_V=2k_{\rm B}$ in Eq. \eqref{cv2} and to $c\tau\approx 1$ \AA\ close to the interatomic separation (UV cutoff in condensed matter) marking the shortest wavelength in the system featuring in Eqs. \eqref{kgap1}, \eqref{harmo} and \eqref{supe1}.

The inversion point in the path-independent ``c''-transition serves as an unambiguous way to separate the liquidlike state at the top part of the graph from the gaslike state at the bottom part. As expected, the line of inversion points on the phase diagram lies close to the FL determined from the dynamical VAF criterion \cite{Cockrell2021}.

From the point of view of aiding the development of the theory of the supercritical state, these results demonstrate a universal interrelation between thermodynamics and dynamics and demonstrate the importance of the quantity $c\tau$ in setting $c_V$ and the loss of degrees of freedom due to the disappearance of transverse modes in the liquidlike state and of longitudinal modes in the gaslike state \cite{Trachenko2016}.

The collapse of all curves up to the key value $c_V=2k_{\rm B}$ has two further implications. First, it suggests that the inversion point $c_V=2k_{\rm B}$ \eqref{cv2} is indeed a special point on the phase diagram as discussed earlier in this section.

Second, if a thermodynamic property has a wide crossover, the behavior of different properties strongly depends on the path taken on the phase diagram. On the other hand, the observed collapse of all paths at the special inversion point close to $c_V=2k_{\rm B}$ and $c \tau = 1$ \AA\ indicates either a sharp crossover or a dynamically driven phase transition related to the ``c"-transition between liquidlike and gaslike states. By sharp we mean more abrupt that transitions observed over the FL in the experimental section II, possibly involving discontinuities in higher-order thermodynamic derivatives. Although a crossover of $c_V$ was detected in molecular dynamics simulations using several fairly sensitive analysis methods \cite{Wang2019}, no $c_V$ anomalies characteristic of a phase transition were seen at temperatures and pressures corresponding to the inversion point within the uncertainty set by fluctuations \cite{Cockrell2021}. Experimental data \cite{nist} do not indicate anomalies close to the FL either, however they are often based on extrapolating the equation of state and interpolating the data. Hence we can not exclude a weak thermodynamic phase transition, similar to a percolation transition, or a higher-order phase transition seen in higher derivatives of thermodynamic functions. This leaves an important question open: does the FL correspond to a crossover or a thermodynamic phase transition, perhaps of higher order? Answering this question may require high-quality modelling data and a new theory of the transition involving novel ideas.

\subsection{Implications for astrophysics and planetary science}

In this section we discuss the implications of the transition at the FL in hydrogen, relevant for astrophysics and planetary sciences. Hydrogen is the most abundant element in the Universe, and has been the subject of the cross-disciplinary research including in condensed matter physics, astronomy and astrophysics. Hydrogen has rich and non-trivial behavior, particularly at high pressure and temperature where it exists in gas giants such as Jupiter and Saturn as well as in hot gaseous exoplanets and brown dwarfs \cite{Helled2020,Guillot1999,McMahon2012}. This is witnessed by advanced modeling and cutting-edge experimental compression techniques as well as space probes to understand the main physical mechanisms at operation in gas giants \cite{Helled2020,Guillot1999,McMahon2012,Redmer2009,Garcia-Melendo2013,Leconte2013,Nettelmann2008,Holst2008,Morbidelli2008,Anderson2007,Stevenson2006,Guillot2005,Gregoryanz2003,Mayer2002,Guillot2002,Chabrier2000}.

In gas giants such as Jupiter, Saturn, exoplanets, and brown dwarfs, molecular hydrogen is supercritical. This fact has been understood according to the earlier view of the supercritical state: moving along any path on a pressure and temperature phase diagram above the critical point was not considered to involve marked changes of properties which were thought to vary in a featureless way. In gas giants, this has been considered \cite{Guillot1999a} to be the case up to high 100-200 GPa pressures where hydrogen fluid dissociates and metallizes \cite{McMahon2012}, and has served as an important starting point of advance modelling and theory \cite{Helled2020,Guillot1999,Redmer2009,Garcia-Melendo2013,Leconte2013,Nettelmann2008,Holst2008,Morbidelli2008,Anderson2007,Stevenson2006,Guillot2005,Mayer2002,Guillot2002,Chabrier2000,Guillot1999a}.

One basic question is related to the supercritical nature of molecular hydrogen, namely whether and how a boundary between the planet's interior and exterior (atmosphere) can be defined? Unlike in terrestrial-type planets such as Earth and Venus where the boundary is clear, the boundary in gas giants is considered as conditional only because the supercritical state was viewed as physically homogeneous and smooth. For practical purposes, the boundary between the interior and the atmosphere is taken at the pressure of the Earth atmosphere of 1 bar \cite{Guillot1999,Guillot1999a}. This gives the radius of about 70,000 km for Jupiter and 57,000 km for Saturn, in approximate agreement with their optical sizes.

\begin{figure*}
\begin{center}
{\scalebox{0.6}{\includegraphics{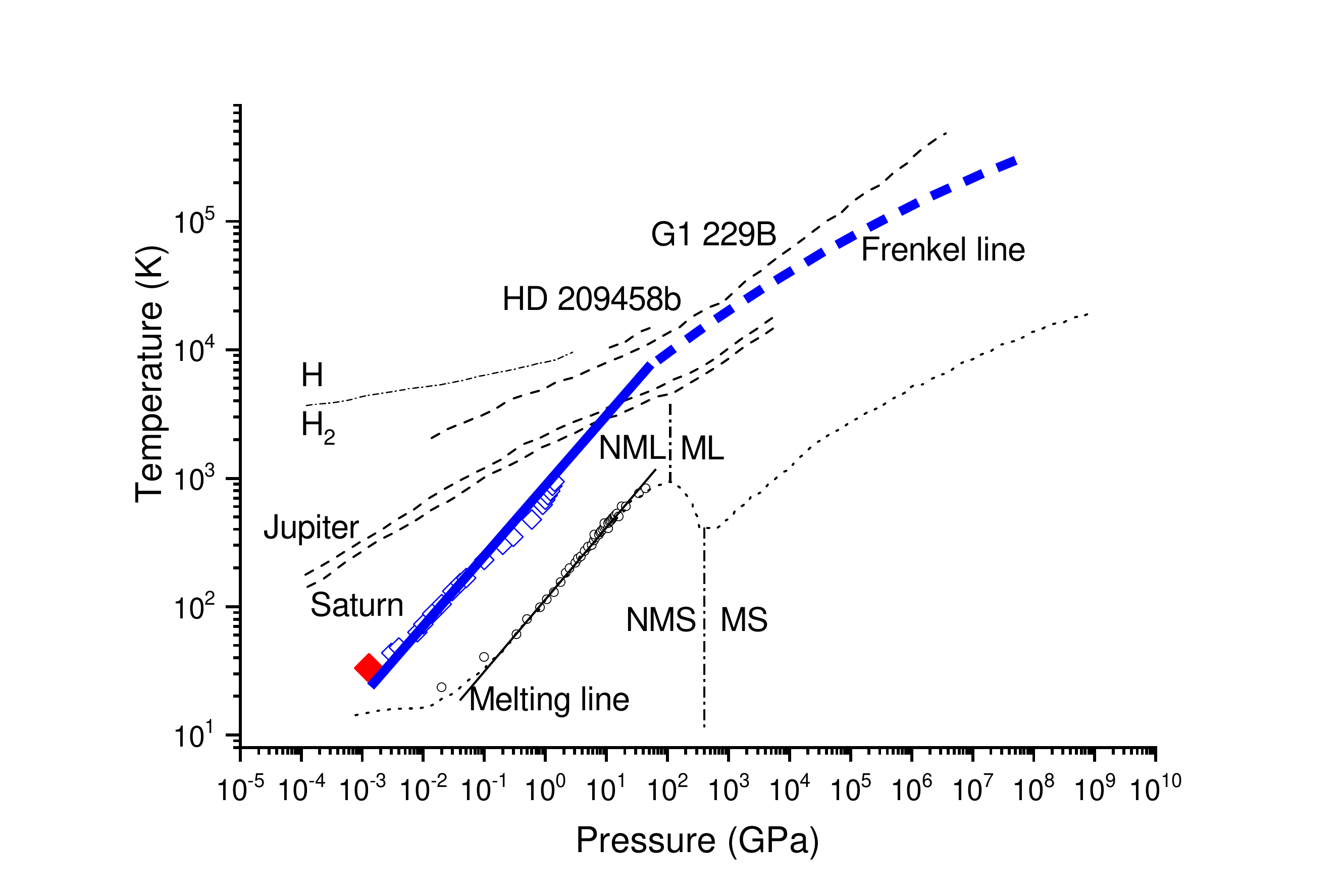}}}
\end{center}
\caption{The adiabats and the Frenkel line in gas giants. Black dashed lines show adiabats inside Jupiter, Saturn, exoplanet HD 209458b and brown dwarf G1 229B \cite{McMahon2012}. Empty black bullets show the melting line of H$_2$ \cite{Gregoryanz2003}, and the straight line showing the linear range. Dotted line shows the melting line \cite{McMahon2012}. Open blue diamonds correspond to pressure and temperature at the minima of the speed of sound \cite{nist}, and solid red diamond shows the critical point of H$_2$. The thick blue line shows the Frenkel line starting just below the critical point and running parallel to the melting line in the range 0.5--50 GPa. The dashed red line shows a tentative extrapolation of the Frenkel line at higher pressure based on existing melting data. Dash-dotted line at low pressure separates molecular hydrogen H$_2$ and dissociated hydrogen H \cite{McMahon2012}. Two vertical dash-dotted lines separate non-metallic solid (NMS) and metallic solid (MS) at low temperature, and non-metallic liquid (NML) and metallic liquid (ML) at high temperature \cite{McMahon2012}. Adapted from Ref. \cite{Trachenko2014}.}
\label{hydrogen}
\end{figure*}

It is interesting to understand whether the effects related to the Frenkel line operate in gas giants. This depends on whether the Frenkel line in hydrogen crosses the adiabat reflecting pressure and temperature conditions in these planets. We use two criteria and data sets to plot the Frenkel line \cite{Trachenko2014}. First, we use the temperature dependence of the isobaric speed of sound measured in supercritical H$_2$ from the NIST database \cite{nist}. The NIST database does not extend to temperatures and pressures high enough as in gas giants, and therefore we use the second criterion: the Frenkel line starts slightly below the critical point and runs parallel to the melting line in the log-log plot \cite{Brazhkin2012,Brazhkin2013}. The parallelity follows from the scaling argument: starting from GPa pressure, the intermolecular interaction is largely reduced to its repulsive part. For the repulsive potential $U\propto\frac{1}{r^n}$, the scaling of pressure and temperature operates: system properties depend on the combination of $TP^\gamma$ only, where $\gamma$ is related to $n$. Consequently, $TP^\gamma= $const holds on all ($P$,$T$) lines where the dynamics of particles changes qualitatively, including the melting line and the Frenkel line. This implies that the Frenkel and melting lines are parallel to each other in the double-logarithmic plot (at low pressure, interactions are not reduced to simple repulsive laws, and the parallelity between the two lines is approximate). We draw the Frenkel line parallel to the melting line in the log-log plot (the melting line is straight in the log-log plot in approximately 0.5--50 GPa range) in Figure \ref{hydrogen}. We observe that the Frenkel line constructed parallel to the melting line is close to the minima of the speed of sound, serving as a consistency check.

We note that extending the Frenkel line above 100 GPa can be done approximately only due to uncertainties of locating the melting line and nonmetal-metal transition in liquid and solid phases of hydrogen \cite{McMahon2012}. For this reason, we draw the Frenkel line at high pressure as a dashed line parallel to the melting line at ultrahigh pressures where hydrogen is monatomic and metallic.

We observe that the Frenkel line crosses the adiabats of Saturn and Jupiter at approximately ($P_{\rm F}=10$ GPa, $T_{\rm F}=3000$ K) and ($P_{\rm F}=17$ GPa, $T_{\rm F}=3900$ K), respectively. This implies that supercritical molecular hydrogen in Saturn and Jupiter exists in two physically distinct states: non-rigid gas-like fluid below ($P_{\rm F}$,$T_{\rm F}$) and rigid liquid above ($P_{\rm F}$,$T_{\rm F}$). The non-rigid gas-like fluid exists in the outer part of the planet, and is separated from the rigid liquid state in the planet interior by the Frenkel line. This takes place at pressures well below 100--200 GPa at which metallization and dissociation of hydrogen take place \cite{McMahon2012} (see Figure \ref{hydrogen}). The transition at the FL is therefore not related to the transition to the atomic metallic phase, but to the molecular hydrogen in the supercritical state.

Unlike in Jupiter and Saturn, the FL does not cross the adiabats of larger and hotter objects: exoplanet HD 209458b and brown dwarf G1 229B. Therefore, supercritical molecular hydrogen in these gas giants is always in the non-rigid gas-like fluid state.

The crossover of particle dynamics corresponds to the qualitative changes of diffusion, viscosity and thermal conductivity, as discussed in the earlier section IIIB and illustrated in Figures \ref{nu} and \ref{alpha}. These changes therefore affect flow, convection and heat transport processes that are at the centre of ongoing theory and modeling of gas giants \cite{Redmer2009,Nettelmann2008,Holst2008,Guillot2005,Chabrier2000,Garcia-Melendo2013,Leconte2013,Guillot1999a,Helled2020}. Although the discussion above concerns hydrogen, it is applicable to any supercritical substance due to the generality of the dynamical crossover at the FL.

The dynamical transition at the Frenkel line can also serve as a physically justifiable boundary between the interior and the exterior (or, if appropriate, the atmosphere) of gas giants. Indeed, the outer matter in planets such as Earth and Venus is below the critical point, and the boundary is clearly defined as the boundary between the gas and the liquid or solid, with first-order phase transitions separating these states.
The separation between the interior and exterior in supercritical gas giants is problematic, and is done on the basis of 1 bar pressure equal to the pressure of the Earth's atmosphere. Convenient for the purposes of comparison, this definition is arbitrary from the physical point of view. On the other hand, the Frenkel line is a physically justified boundary applicable to gas giants. As we saw earlier, the character of particle motion qualitatively changes at the FL, with the accompanying changes of all major physical properties from gas-like to liquid-like. These changes are similar to those taking place at the liquid-gas planet-atmosphere boundary in smaller planets such as Earth or Venus except without a first-order phase transition. For example, rigidity, the stability against solid-like shear distortions, exists in the gas giant interior below the FL but not in the exterior above the FL (in solids rigidity is static but in liquids it is dynamical and operates at large frequency or wavevector as discussed earlier). Similarly, on planets such as Earth and Venus, rigidity changes on the surface, and separates planet's interior from the atmosphere: the solid crust has static rigidity and the liquid ocean has dynamical rigidity at high frequency and wavevector, whereas gas in a nonrigid state. Using the relationship between pressure and density \cite{Guillot1999a}, values of $P_{\rm F}$ above and the hydrostatic relationship $P_{\rm F}=g\rho H_{\rm F}$, where $H_{\rm F}$ is the height of the Frenkel line boundary below the current 1 bar ``surface'', $H_{\rm F}$ can be estimated to be ($4000\pm 1000$) km and $(6000\pm 1000$) km in Jupiter and Saturn, respectively, corresponding to 66,000 km from the centre in Jupiter (94\% of the currently used radius) and 51,000 km in Saturn (89\% of the current radius) \cite{Trachenko2014}.

Our final remark is related to another property of hydrogen at high pressure relevant for astrophysics and planetary science, the speed of sound. In section IIIB, we discussed the lower bounds of viscosity and thermal conductivity which can be expressed in terms of fundamental physical constants. It was recently shown \cite{soundbound} that the upper bound on the speed of sound in condensed phases, $v_u$, can be similarly written in terms of fundamental constants as

\begin{equation}
\frac{v_u}{c}=\alpha\left(\frac{m_e}{2m_p}\right)^{\frac{1}{2}}
\label{sound}
\end{equation}
\noindent where $c$ is the speed of light in vacuum, $\alpha$ is the fine-structure constant and $m_e$ and $m_p$ are electron and proton masses.

Eq. \eqref{sound} applies to the lightest element, atomic hydrogen, and gives the upper bound $v_u$ of about 36 km/s, in agreement with the value calculated in the high-pressure phase of solid atomic metallic hydrogen in the pressure range above 250 GPa \cite{soundbound}.

\section{Summary and outlook}

We have reviewed the history of supercritical research and discussed recent experiments stimulated by the FL. The experiments include Ne, N$_2$, CH$_4$, C$_2$H$_6$, CO$_2$ and H$_2$O using X-ray, neutron and Raman scattering techniques and evidence the transition at conditions predicted by the FL. We also reviewed areas where the FL research branched out, including percolation, topology, mode decomposition and quantum path integral simulations. As we survey and chart deep supercritical state, further experiments will be important and reveal new physics in supercritical fluids. This research will benefit from experimental advances in scattering techniques shaping the future of supercritical research.

While we have a fairly good understanding of thermodynamics of supercritical fluids below and above the FL, an important open question is the nature of the transition at the FL itself. Progress in this and related areas awaits novel ideas and new high-quality simulation and experimental data.

\section{Acknowledgements}

We are grateful to EPSRC and RSF (19-12-00111) for support.

\bibliography{collection_2021}

\bibliographystyle{ieeetr}

\end{document}